\documentstyle[11pt,aaspp4]{article}
\def \lta {\mathrel{\vcenter
     {\hbox{$<$}\nointerlineskip\hbox{$\sim$}}}}
\def \gta {\mathrel{\vcenter
     {\hbox{$>$}\nointerlineskip\hbox{$\sim$}}}}
\def \m{\ifmmode M_\odot\else M$_\odot$\fi}
\def \r{\ifmmode R_\odot\else R$_\odot$\fi}

\def\gcm3{g~cm$^{-3}$}
\def\g-s{g~s$^{-1}$}
\def\cm3s{cm$^3$~s$^{-1}$}      
\def\cms{cm~s$^{-1}$}
\def\kms{km~s$^{-1}$}
\def\erg-s{erg~s$^{-1}$}     

\def\beq{\begin{equation}}
\def\eeq{\end{equation}}
\def\gr{$\gamma$-ray}
\def\grs{$\gamma$-rays}
\def\grb{$\gamma$-ray burst}
\def\grbs{$\gamma$-ray bursts}          
\def\0{\parindent=0cm}
\def\5{\parindent=.5cm}
\def\7{\parindent=.7cm}
\def \lta {\mathrel{\vcenter
     {\hbox{$<$}\nointerlineskip\hbox{$\sim$}}}}
\def \gta {\mathrel{\vcenter
     {\hbox{$>$}\nointerlineskip\hbox{$\sim$}}}}

\begin{document}

\title{Asymmetric Supernovae, Pulsars, Magnetars, and Gamma-Ray Bursts}
\author{J. Craig Wheeler $^1$, Insu Yi$^{2, 3}$, Peter H\"oflich $^1$
and Lifan Wang $^1$}
\affil{$^1$Astronomy Department, University of Texas, Austin, Texas 78712;
 wheel@astro.as.utexas.edu, pah@alla.as.utexas.edu, lifan@tao.as.utexas.edu}
\affil{$^2$Korea Institute for Advanced Study, 207-43 Cheongryangri, Dongdaemun,
Seoul, 130-012, Korea; iyi@kias.re.kr}
\affil{$^3$Institute for Advanced Study, Princeton, NJ 08540}

\begin{abstract}

We outline the possible physical processes, associated
timescales, and energetics that could lead to the production of pulsars,
jets, asymmetric supernovae, and weak \grbs\ in routine 
circumstances and to a $10^{16}$ G magnetar and perhaps stronger \grb\
in more extreme circumstances in the collapse of the bare core 
of a massive star.  The production of a LeBlanc-Wilson MHD jet could
provide an asymmetric supernova
and result in a weak \grb\ when the jet accelerates down the stellar
density gradient of a hydrogen-poor photosphere.  
The matter-dominated jet would be formed promptly, 
but requires 5 to 10 s to reach the surface of the progenitor of a Type Ib/c supernova.
During this time, the newly-born neutron star could contract, spin up,
and wind up field lines or turn on an $\alpha-\Omega$ dynamo.  
In addition, the light cylinder
will contract from a radius large compared to the Alfv\'en radius
to a size comparable to that of the neutron star.  This will
disrupt the structure of any organized dipole field and promote
the generation of ultrarelativistic MHD Waves (UMHDW) at high density and
Large Amplitude Electromagnetic Waves (LAEMW) at low density.  
The generation of these waves would be
delayed by the cooling time of the neutron star $\simeq$ 5 to 10 seconds,
but the propagation time is short so the UMHDW could arrive at the surface
at about the same time as the matter jet.  In the density
gradient of the star and the matter jet, 
the intense flux of UMHDW and LAEMW could drive shocks, generate pions
by proton-proton collision, or create electron/positron pairs 
depending on the circumstances.  The UMHDW and LAEMW could influence the 
dynamics of the explosion and might also tend to flow out the rotation axis
to produce a collimated \grb.

\end{abstract}

\keywords{supernovae: general $-$ gamma rays: bursts $-$ pulsars:general $-$
ISM: jets and outflows}

\newpage

\section{Introduction}

Recent evidence has given support for the idea that the core-collapse 
process is intrinsically strongly asymmetric. 
The spectra of Type II and Type Ib/c supernovae are significantly polarized
indicating asymmetric envelopes (M\'endez et al. 1988; Jeffrey
1991; H\"oflich 1991; Trammel et al. 1993; Wang et al. 1996; 
Tran et al. 1997; Leonard et al. 1999).  The degree of polarization  
tends to vary inversely with the mass of the hydrogen envelope
(Wang, Wheeler \& H\"oflich 2000). 
Pulsars are observed with high velocities, up to 1000 \kms\ 
(Strom et al. 1995).  Observations of SN~1987A showed that radioactive 
material rapidly mixed out to the hydrogen-rich layers 
(Lucy 1988; Sunyaev et al. 1987; Tueller et al. 1991).
Cas A shows rapidly moving oxygen-rich
matter outside the nominal boundary of the remnant  (Fesen \&
Gunderson, 1996) and  evidence for two oppositely directed jets of
high-velocity material (Reed, Hester, \& Winkler 1999).
High velocity ``bullets" of matter have been observed in the Vela
supernova remnant (Taylor et al. 1993).            
Other evidence shows that soft gamma-ray repeaters
arise in very strongly magnetized neutron stars, ``magnetars,"
with dipole fields in the range $10^{15}$ G (Kouveliotou et al. 1998;
Thompson et al. 1999).  
Theoretical models have shown that matter-dominated jets can cause 
supernova explosions (Khokhlov et al. 1999) and that a strongly 
asymmetric explosion can account for 
the outward mixing of $^{56}$Ni in SN~1987A
(Nagataki 1999).  In addition, 
much attention has recently been paid to the issue of collimation of
\grbs\ and the associated affect on energetics and
observable properties in the context of high-redshift events 
(Kulkarni et al. 1999; Rhoads, 1999; Sari, Piran \& Halpern 1999;
Lamb 1999; Chevalier \& Li 1999a,b) and the 
proposed correspondence of SN~1998bw with a much
lower energy event (Galama et al. 1998).
In this paper, we explore the physics that might unite all these
areas.

The discovery of optical afterglows of \grbs\ has raised the
estimates of the maximum isotropic \gr\ emission to unprecedented values,
$\sim$ $3\times10^{54}$ ergs for GRB 990123 (Kulkarni, et al. 1999).
This has brought a new focus on the liklihood that the prompt
\grbs\ and the afterglows are collimated to various extents as
well as Doppler boosted and ``beamed"  (Rhoades 1997, 1999;
Sari, Piran \& Halpern 1999; Kulkarni et al. 1999; Stanek et al. 1999;
Harrison et al.  1999).  Although the evidence is still preliminary,
collimation factors of $\Delta\Omega/4\pi\lta 0.01$ have been
derived from the decline in some afterglow light curves.  
Wang \& Wheeler (1998), Nakamura (1998) and
Cen (1998) have pointed out that if the collimation were strong
enough, the energetics might be provided by supernova-like energies.
SN~1998bw and GRB~980425 provided a different perspective by
suggesting that some \grbs\ are directly associated with some
supernovae (Galama et al. 1998; see also Bloom et al. 1999;
Reichart 1999; Germany et al. 1999; Galama et al. 1999; Wheeler 1999).  
If this association is correct,  
the ``isotropic"  \gr\ energy of GRB~980425 is only $\simeq10^{48}$ ergs.
Some models of SN~1998bw invoked especially large kinetic energies,
in excess of $10^{52}$ ergs in spherically-symmetric models, to 
account for the bright light curve and high velocities (Iwamoto et al.
1998; Woosely, Eastman \& Schmidt 1998), and others took note of
the measured polarization and chemical structure
to suggest that strongly asymmetric
models could account for the observations with more ``normal"
energies (H\"oflich, Wheeler \& Wang 1999; Danziger et al. 1999).  
It is not clear that either class of models can naturally
produce a \grb\ of $\sim10^{48}$ ergs.  The former models
have been referred to as ``hypernova" models, although that
term was originally introduced (Paczy\'nski 1998) to mean the
generic high energy events associated with the afterglows of
the classical \grbs\ at large distance.  

Many of the popular models for creating \grbs\ are based on 
binary neutron stars (Paczy\'nski 1986) 
or accretion onto black holes (Paczy\'nski 1991; Woosley 1993).
The latter, in particular, are popular because there is no
limit, in principle, to the mass of the black hole and hence,
again in principle, the energy that can be extracted.  All the
models struggle with the mechanism for turning the large
energies into \grs. There is no reason why the energy flux
from a black hole source cannot be collimated, as witness the
jets from active galactic nuclei and some binary black hole sources.
This could reduce the energy requirements even in black hole models.
Some models explicitly include this collimation (MacFadyen \& Woosely  
1999).  We note that even the superluminal jets from AGN and
blazars have a maximum bulk Lorentz factor of about 10, while
the \grbs\ require $\Gamma\gta 100$ (Krolik \& Pier 1991;
Baring \& Harding 1997:  Piran 1999,
and references therein). 
This may suggest that a
qualitatively different mechanism is needed to generate the
cosmic \grbs\ despite the attractive possibilities for black hole models.  

Here we will attempt to see how far a more conservative 
model can go, both to produce asymmetric supernovae and perhaps to
generate \grbs, by considering the effects of a newly-born pulsar.
This is not a new idea (Ostriker \& Gunn 1971; Bisnovatyi-Kogan 1971;
Bisnovatyi-Kogan et al. 1975),
but it is worth reconsidering in the context of the polarization
of core-collapse supernovae, 
the growing evidence for ``magnetars," 
the growing understanding of \grbs\ and their afterglows, the likely
association of SN~1998bw with GRB~980425, and the interesting 
possibility that supernovae can be induced by energetic jets
arising from their cores.  All these processes
demand or suggest strong asymmetries. Questions arise as to whether
these issues are related, whether \grbs\ of observed properties
could be produced in this context, whether there are more than one type
of \grb\ mechanism, and whether supernovae may, in some circumstances,
produce the \grbs\ seen at high redshift.         
It is very plausible that the formation of a neutron star will
engender collimated flow, including Large Amplitude
Electromagnetic Waves (LAEMW: Usov, 1992, 1994; Blackman \& Yi 1998).  
Here we will show that
it is plausible that the properties of SN~1998bw, including a weak
\grb, may be generated by the acceleration of a collimated shock
down a density gradient and possible that a \grb\ visible
at cosmic distances could be produced under some circumstances.
Certain aspects of our model are similar to those of Usov (1992, 1994),
but he considered accretion-induced collapse and we explicitly 
investigate the context of a core collapse in the core of a massive
star.  Some of the ideas we explore here are also presented
by Nakamura (1998).

The purpose of this paper is to outline the basic time scales,
energetics, and relevant physical processes in order to define areas
that need more quantitative work.  Key length
and time scales and a summary of the core collapse ambiance
are given in \S 2 and \S 3,
respectively.  The physics of the proto-neutron star phase including
the generation of the magnetic field and associated torquing of
the surrounding plasma is presented in \S 4.  The effect of an
axial jet associated with the collapse process is outlined in
\S 5.  The important phase when the neutron star cools, contracts,
and spins more rapidly is discussed in \S 6, and the manner in
which the energy associated with matter and radiation jets
could be propagated outward is given in \S 7.  
Discussion and conclusions are given in \S 8.

\section{Basic Length and Time Scales}

In the discussion to follow there are a number of key length and
time scales.  Among these are the radius of the star, $R_{\rm star}$
$\simeq 10^{8}$ km for a red supergiant and $R_{\rm He}$ 
$\simeq 2\times 10^{5}$ km
for a helium core.  The inner iron core that collapses to form
a neutron star has a typical radius, $R_{\rm Fe}\simeq 4\times10^{3}$ km.    
The dynamical or sound crossing time is
\begin{equation}
\label{dynamictime} 
\tau_{\rm dyn}\simeq 20 {\rm s} \frac{R_{5}^{3/2}}{(M/\m)^{1/2}},
\end{equation}
where $R_{5}$ is the radius in units of $10^{5}$ km,
and the light travel time is 
\begin{equation}
\label{lightcross} 
\tau_{{\rm light}}~=~\frac{R}{c}\simeq 0.3~{\rm s}~R_{5}.
\end{equation}       
The dynamical time to emerge from a red giant or even a light crossing time, 
from $10^3$ to $10^5$ s, is not likely to be associated with
observed \grbs.  For a stripped helium core, the progenitor of
a Type Ib or Type Ic supernova, a range of timescales
from 1 to 10 s is possible, in the absence of Doppler boosting.
Note that the breakout of a collimated jet might involve an area
considerably less than the surface area of the helium core and
an appropriately smaller time scale.
Other relevant length scales are the radius of the light cylinder, 
\begin{equation}
\label{lcylinder} 
R_{LC}~=~\frac{c}{\Omega}~=~\frac{cP}{2\pi} \simeq~50~{\rm km}
\left(\frac{P}{{\rm ms}}\right),
\end{equation}
where $\Omega$ is the rotational frequency and P the rotational period
of the neutron star, and the Alfv\'en radius at
which magnetic pressure is balanced by the ram pressure, e.g.,
\begin{equation} 
\label{magpressure} 
\frac{1}{2}\rho v^2\simeq\frac{1}{8\pi}B^2,
\end{equation}          
which, with $B\simeq B_{NS}\left(\frac{R}{R_{NS}}\right)^{-3}$, is 
\begin{equation}   
\label{alfven} 
R_A\simeq 3.0 R_{NS}
\left(\frac{B}{10^{14}~{\rm G}}\right)^{1/3}
\left(\frac{\rho}{10^8~{\rm g~cm^{-3}}}\right)^{-1/6}
\left(\frac{v}{10^8~{\rm cm~s^{-1}}}\right)^{-1/3}.
\end{equation} 

During the thermal contraction of the proto-neutron star, 
P, $R_{LC}$, and $R_A$ will all change significantly.  
For illustration, we assume that the rotation period decreases from 
that of the proto-neutron star, $P_{PNS} \simeq 25$ ms,
to that of the  neutron star, $P_{NS} \simeq 1$ ms.  
As a result, the light cylinder will contract
from $R_{LC} \simeq 10^3$ km, beyond the radius of the stalled shock
and comparable in size to the original iron core, to $R_{LC} \simeq 50$ km,
comparable to the radius of the neutron star and well within the
stalled shock.  The light cylinder will 
contract from beyond $R_{A}$ to significantly less than $R_{A}$.
This has critical implications for angular momentum transport and
the generation of radiation.

\section{Context of Core Collapse}

We know that pulsars arise from core-collapse.  We know that some
core collapse events are directly associated with supernovae, e.g.,
the Crab Nebula, SN~1987A.  Circumstantial evidence suggests that
supernovae of Type II and Ib/c result from core collapse based
on correlations with spiral arms and H II regions in spiral galaxies and 
on the nebular spectra that are consistent with the cores of massive
stars.  Polarization data suggests that all these
events are significantly asymmetric and that the asymmetry is
stronger for explosions with smaller hydrogen envelopes
(Wang, Wheeler \& H\"oflich 1997; Wang \& Wheeler 1998; 
H\"oflich, Wheeler \& Wang 1999; Leonard et al. 1999;
Wang, Wheeler \&  H\"oflich 2000).  Explosions induced by
jets from within the inner core can plausibly account for
this asymmetry (Khokhlov et al. 1999).  The imprint of such
jets and the possibility to make \grbs\ is largest in massive
stars from which the hydrogen envelope has been stripped by
winds or binary mass transfer.  In the following we will
thus concentrate on core collapse explosions and neutron star
formation in hydrogen and helium-deficient
Type Ib and Type Ic supernovae.  Such supernova progenitors must be 
surrounded by substantial mass from the previous mass loss stages,
and this will affect the production and evolution of 
jets and \grbs. 

Here we will adopt a generic
picture of core collapse in an intermediate mass star
of main sequence mass $\simeq$ 
15 - 25 \m\ with length scales as given in \S 2.  
The iron core undergoes instability and collapses in a time of
$\simeq$ 1 s.  The collapse involves a homologous collapse of the
iron core in which the density increases monotonically inward.  
The outer parts of the stellar core composed of silicon, oxygen, carbon,
etc, have lower densities and longer free-fall times.  Over the times
of interest here, $\simeq 10$ s, these outer layers will ``hover" as the
collapsing iron core succeeds or fails to generate an explosion.

After core bounce, a proto-neutron star (PNS) forms
with a radius of $R_{PNS}\simeq 50$ km and a shock is formed that
stalls at a radius of $R_{sh}\simeq 200$ km.  Over a cooling time
$t_{cool}\simeq$ 5 to 10 s, the proto-neutron star de-leptonizes by 
neutrino emission, cools, and contracts to form the final neutron
star structure (Burrows \& Lattimer 1986).  If the neutron star
is rotating and magnetized, this cooling phase will be associated
with contraction, spin-up, and amplification of the magnetic
field.  These changes can significantly alter the physics associated
with the neutron star and its interaction with its surroundings
and hence the explosion process itself.

\section{The Proto-Neutron Star Phase}

Right after collapse, the hot proto-neutron star undergoes 
neutrino-driven convection with the characteristic convective
overturn time scale $\tau_{\rm conv}\simeq{\rm 1~ms}~F_{39}^{-1/3}$ where
$F_{39}\equiv F/10^{39}$ \erg-s is the neutrino flux at
the base of the convection (Duncan \& Thompson 1992). The hot 
proto-neutron star has a radius $\simeq$~50 km that gradually shrinks
on the cooling time scale $\simeq$~ 5 to 10 s.  This time scale is
coincidently, but significantly, similar to the sound crossing
time of the helium core, and hence about the time required for
a supersonic (but not relativistic) jet to penetrate the
outer mantle.  There are thus two potential mechanisms to 
create a \grb\ about 10 s after collapse.  A \grb\ could be generated as the
jet accelerates down the density gradient at the boundary of the core.
Alternatively, a strong flow of Poynting flux might
be generated on the neutron-star cooling time as the neutron
star spins up.  This might lead to an alternate mechanism of \grb\
formation on a similar timescale (Nakamura 1998).  
These mechanisms for generating
\grbs\ will be discussed in \S 7.

It is hard to estimate the spin
period of the proto-neutron star when it first forms.  We
assume that any \grb\ stimulated by Poynting from the neutron star
will occur near the end of the cooling, contraction
stage when the neutron star spins with a period $P_{NS}\simeq1$ ms.  
As we will see below, a spin period approaching $P_{NS}\simeq1$ ms is
necessary (if not sufficient) to power a classical \grb.  
A slower spin or shock breakout associated with a jet might generate
the type of \grb\ observed in SN~1998bw/GRB~980425 as we will
also show below.   
For purposes of illustration, we then adopt a spin period of the 
proto-neutron star to be that which will lead to a spin
period of about 1 ms after cooling and contraction.  
Assuming angular momentum conservation during contraction
of the original proto-neutron star, that 
$I_{PNS}\equiv kMR_{PNS}^2$ with $k=$ const $\simeq\frac{2}{5}$, and that 
$R_{NS}$ shrinks from 50 km to 10 km during the deleptonization phase,
we take:
\begin{equation}
\label{PNSperiod} 
P_{PNS}\simeq{\rm1 ms}\left(\frac{50~{\rm km}}{10~{\rm km}}\right)^2
\simeq 25~{\rm ms}.
\end{equation}

We note that this spin rate for the proto-neutron star implies
a rather rapid rate of spin for the original iron core.
Conservation of angular momentum (assuming conservation
of mass and that the form factor for the moment of inertia
is unchanged) gives
\begin{equation}
\label{ironperiod}
\Omega_{Fe}\simeq\frac{M_{PNS}}{M_{Fe}}\left(\frac{R_{PNS}}{R_{Fe}}\right)^2
\Omega_{PNS}\simeq{\rm 0.04 s},
\end{equation}
or a period of about 2.6 minutes.  This represents a rotational energy of about 
$10^{47}$ erg, an amount that is significant, but still much smaller than
the binding energy of the iron core, $\simeq10^{51}$ erg.  If the total
helium core is in solid body rotation with the same period of about
3 minutes, then for a mass of 5 \m\ and a radius of $2\times10^{10}$ cm,
the rotational energy would be about $10^{51}$ erg.  This is clearly
a demanding criterion.  The specific parameters we adopt here in 
equation (\ref{PNSperiod}) and below are probably only reasonable if
there is substantial differential rotation between the iron core and
the surrounding star.

The moment of inertia of the proto-neutron star 
is $I_{PNS}\simeq 3\times 10^{46}$ c.g.s.
With the fiducial period from equation (\ref{PNSperiod}) the
rotational energy of the proto-neutron star is then:
\begin{equation}
\label{littleErot} 
E_{\rm rot,PNS}\simeq\frac{1}{2}I_{PNS}\Omega_{PNS}^2
\simeq 9\times10^{50}\ {\rm erg}
\left(\frac{M_{NS}}{1.5 {\m}}\right)
\left(\frac{\Omega_{PNS}}{250~{\rm s^{-1}}}\right)^2
\left(\frac{R_{PNS}}{50~{\rm km}}\right)^2.           
\end{equation}
This is a substantial energy, but since only a fraction of it could
be tapped it is not clear that this energy could power a supernova
never mind a classical \grb.  If the rotational period of the proto-neutron 
star is longer, this is even more true. The fraction of this rotational
energy that can be tapped depends on the behavior of the magnetic
field, and we turn to that subject. 

\subsection{Magnetic Field Amplification}
 
The magnetic field of the proto-neutron star is uncertain.   If the
pre-collapse core has a field strength comparable to that
of a magnetized white dwarf, 
$\simeq10^6$ G, then a field of $\simeq10^{12}$ G could
arise from flux-freezing.  For $P_{PNS}\simeq$ 25 ms 
(equation \ref{PNSperiod}) and $\tau_{\rm conv}\simeq$ 1 ms,
the Rossby Number $R_0\simeq P_{PNS}/\tau_{\rm conv}\simeq 25\gg1$ 
is too large to allow an $\alpha-\Omega$ type dynamo to
operate.  An alternative magnetic field
amplification mechanism is linear amplification (Meier et al. 1976,
Klu\'zniak \& Ruderman 1998).  In this process,
the differentially rotating neutron star could wrap the poloidal
seed field into strong toroidal fields which then emerge from the star
through buoyancy.  After $n_\phi$ revolutions of the neutron star, 
the initial seed (poloidal) field is wrapped and amplified to
produce a toroidal field
\begin{equation}
\label{Bphi} 
B_\phi\simeq2\pi n_\phi B_p,
\end{equation}
where $B_p$ is the initial seed poloidal field.  Buoyancy will operate 
to expel the field if the amplified final field $B_f$ satisfies
\begin{equation}
\label{buoyancy} 
\frac{B_f^2}{8\pi}\simeq f_{B} \rho c_s^2,
\end{equation}
where $f_{B}\simeq0.01$ is the fractional difference in density 
between the rising flux tube elements and the stellar material. 
Assuming sound speed $c_s\simeq$ c/3 and density 
$\rho\simeq10^{13}$ \gcm3, one finds
\begin{equation}
\label{Bbuoyancy}
B_f\simeq2\times10^{16}~{\rm G}~f_{B-2}^{1/2}\rho_{13}^{1/2},
\end{equation}
where $f_{B-2} = f_B/0.01$ and $\rho_{13}=\rho~/10^{13}$ \gcm3.
Assuming that the magnetic flux tube (a torus) occupies a volume of
$V_B/V_{PNS}\simeq0.1$ where $V_{PNS}\simeq\frac{4\pi}{3}R_{PNS}^3$,
the energy contained in the magnetic flux tubes at the
buoyancy limit is estimated to be:
\begin{eqnarray}
\label{Ebuoyancy} 
E_B&\simeq&0.1\times V_{PNS}\times\frac{B_f^2}{4\pi} \\
&\simeq&1.6\times10^{51}\ {\rm erg}.\nonumber
\end{eqnarray}

This magnetic energy is ejected from the 
neutron star in the rising magnetic flux tubes.
This energy is comparable to the 
proto-neutron star rotation energy (equation \ref{littleErot}).  For the
adopted parameters, the proto-neutron star would have to lose substantial
rotational energy to the magnetic field before the field would
float on dynamical time scales. This magnetic energy is still likely to
escape from the proto-neutron star, but the details may be complex
and involve the subsequent contraction of the neutron star. 

The number of revolutions to reach $B_f \simeq 2\times10^{16}$ G is:
\begin{equation}
\label{nwrap} 
n_f=\frac{B_f}{B_0}\frac{1}{2\pi}\simeq3\times10^3
\left(\frac{B_0}{10^{12}~{\rm G}}\right)^{-1}.
\end{equation}
For $B_0\simeq10^{12}~{\rm G}$, the amplification time scale before the
field is expelled by buoyancy is thus
\begin{equation}
\label{twrap} 
t_f\simeq n_fP_{PNS}\simeq25~{\rm ms}\times3\times10^3\sim 75~{\rm s},
\end{equation}
at the beginning of the contraction of the proto-neutron star.  
As the neutron star contracts, it spins up.
The timescale for the linear field amplification, $t_f$, 
will decrease and the rotational energy of the neutron star will
increase (cf \S 6).  By the time the neutron star is spinning
with a period of 1 ms, $t_f$ would be substantially less than
the cooling, contraction time of $\simeq10$ s.  
Thus sometime during the contraction a dipole-type strong field is
expected to be produced once the
$\simeq 10^{16}$~G field emerges from the surface. A
dipole field strength of $\simeq 10^{14}$~G is expected from the 
random sum of flux tubes of $\simeq 10^{16}$~G (Duncan \& Thompson 1992;
Thompson \& Duncan 1993).  
It is likely that during the contraction phase a substantial fraction of
$E_{rot,PNS}\simeq E_{B,PNS}\simeq 10^{51}$ erg 
would be extracted from the neutron star.  

\subsection{Energy Transport}

We now consider the possible energy transport 
mechanisms from the neutron star to the outer stellar envelope.  
For $P\simeq25$ ms, the radius of the light cylinder is
(cf. equation \ref{lcylinder}) $R_{LC,PNS}\simeq10^3$ km,
comparable to the initial radius of the iron core and much
larger than the proto-neutron star.  For $R<R_{LC,PNS}$,
a rotating dipole field could
exert a magnetic torque on the plasma around the star.  
Typical collapse calculations (Burrows, Hayes \& Fryxell 1996;
Mezzacappa, et al. 1997) give for the density in the
vicinity of the standing shock $\rho\simeq10^8$ \gcm3 with
a pre-shock velocity of $v\simeq10^8$ \cms.  Using those
as fiducial values, the characteristic 
Alfv\'en radius for the proto-neutron star is
(equation \ref{alfven}):
\begin{equation}
\label{alfvenPNS} 
R_A\simeq~150\ {\rm km}
\left(\frac{R_{PNS}}{50~{\rm km}}\right)
\left(\frac{B_{PNS}}{10^{14} ~{\rm G}}\right)^{1/3}
\left(\frac{\rho}{10^8~{\rm g~cm^{-3}}}\right)^{-1/6}
\left(\frac{v}{10^8~{\rm cm~s^{-1}}}\right)^{-1/3}.
\end{equation}
Equation (\ref{alfvenPNS}) implies that for 
$B_{PNS}$ substantially less than $10^{14}$~G the Alfv\'en radius
would be less than the radius of the proto-neutron star and hence
that the magnetic torque would be ineffective. 
For $B_{PNS}\gta10^{14}$~G, the torque could have a significant effect.
Equation (\ref{alfvenPNS}) shows that $R_A\ll R_{LC}\simeq 1000$ km,
so that a dipole field could be maintained over a significant volume. 
For the fiducial values chosen in equation (\ref{alfvenPNS}), $R_A$ 
is interestingly close to the stalled shock radius $\simeq$ 200 km,
thus justifying the choice of density and velocity, to which
$R_A$ is not, in any case, very sensitive.

There are two relevant timescales pertinent to the action of
the torque.  One is the time to deflect infalling matter from
radial infall to substantially azimuthal flow.  The second is
the time for this torque to spin down the proto-neutron star and
hence to deposit energy in the infalling matter.
The time for the torque to deflect infalling matter is
approximately the momentum per unit volume of the infalling
matter times an appropriate lever arm, $R_A$, divided by the
torque per until volume, $\simeq B_z(R)B_{\phi}(R)\simeq B^2(R)$
(cf. Shapiro \& Teukolsky 1983).  This effect of the torque 
thus operates on a time scale
\begin{eqnarray}
\label{torquetime1} 
t_{\rm tor,1}&\simeq&\frac{\rho R_A v}{B^2}\simeq\frac{\rho R_A v}{4\pi\rho
v^2}\simeq\frac{1}{4\pi}\frac{R_A}{v},\\
&\simeq&0.01 {\rm s}\left(\frac{R_{PNS}}{50~{\rm km}}\right)   
\left(\frac{B_{PNS}}{10^{14}~{\rm G}}\right)^{1/3}
\left(\frac{\rho}{10^8~{\rm g~cm^{-3}}}\right)^{-1/6}
\left(\frac{v}{10^8~{\rm cm~s^{-1}}}\right)^{-4/3}.\nonumber
\end{eqnarray}
This time is sufficiently short that the torque could substantially
alter the flow of matter, preventing the radial infall interior to
the shock that is common to all spherically symmetric models of
core collapse.  Multi-dimensional models that produce non-radial
circulation flows in the matter beneath the standing shock would
also be substantially affected for conditions similar to those 
reflected in equation (\ref{torquetime1}). 

The time scale to extract energy from the proto-neutron star and
deposit that energy in the surrounding plasma can be estimated
by equating the torque, N, on the proto-neutron star to that on
the gas,
\begin{equation}
\label{torque}
N = I{\dot \Omega} \simeq \int\nolimits^{\infty}_{R_A}R^2B_z(R)B_\phi(R)dR
\simeq R_A^3 B(R_A)^2.
\end{equation}
Defining the timescale appropriate to this action of the torque to
be $t_{tor,2}=\Omega/{\dot \Omega}$ and using equation (\ref{torque})
gives:
\begin{eqnarray}
\label{torquetime2}   
t_{\rm tor,2}&\simeq&\frac{1}{5\sqrt{\pi}}\frac{\Omega_{PNS} M_{NS}}
{R_{PNS}B_{PNS}\rho^{1/2}v},\\
&\simeq&170~{\rm s}
\left(\frac{\Omega_{PNS}}{250~{\rm s^{-1}}}\right)
\left(\frac{M_{NS}}{1.5 \m}\right)
\left(\frac{R_{PNS}}{50~{\rm km}}\right)
\left(\frac{B_{PNS}}{10^{14}~{\rm G}}\right)^{-1}
\left(\frac{\rho}{10^8~{\rm gm~cm^{-3}}}\right)^{-1/2}
\left(\frac{v}{10^8~{\rm cm~s^{-1}}}\right)^{-1}.\nonumber
\end{eqnarray}  
The rate of energy deposition in the plasma by this torque is
thus:
\begin{eqnarray}
\label{torqueenergy}
L_{tor}&=&I\Omega{\dot \Omega} 
\simeq 2\sqrt{\pi}\Omega R_{PNS}^3B_{PNS}\rho^{1/2}v,\\
&\simeq&1\times10^{49}~{\rm erg~s^{-1}}
\left(\frac{\Omega_{PNS}}{250~{\rm s^{-1}}}\right)
\left(\frac{R_{PNS}}{50~{\rm km}}\right)^3
\left(\frac{B_{PNS}}{10^{14}~{\rm G}}\right)
\left(\frac{\rho}{10^8~{\rm g~cm^{-3}}}\right)^{1/2}
\left(\frac{v}{10^8~{\rm cm~s^{-1}}}\right).\nonumber
\end{eqnarray}     
Note that since the torque is independent of $\Omega$, the angular
frequency will tend to decrease linearly.	
The timescale $t_{tor,2}$ is sufficiently long and the associated
energy deposition rate in equation (\ref{torqueenergy}) is
sufficiently small that it is unlikely that this energy deposition
will directly contribute to any supernova explosion.
On the other hand, the selective deposition of this energy very
near the stalled shock might have a leveraging effect
belied by the small deposition rate that could help to
reinvigorate a stalled shock in conjunction with other effects
such as neutrino deposition.
 
\section{The Effect of an Axial Jet}

If, as in LeBlanc \& Wilson (1970; see also M\"uller \& Hillebrandt 1979;
Symbalisty 1984), an MHD jet is formed 
during the collapse phase, the maximum jet power is estimated to be 
$E_{\rm rot}/\tau_{dyn}\simeq10^{51}$ erg/1 s
$\simeq10^{51}$ \erg-s.  Although the details of the jet dynamics and
energetics are not known, roughly $\lta10^{51}$ \erg-s of power
could be directed along the jet axis.  
The LeBlanc \& Wilson calculation was criticized by Meier et al. (1976)
as requiring extreme parameters of the progentor star.  These issues
need to be re-examined in the current context, but we note several
things about the Meier et al. analysis.  They argue that the MHD
axial flow found by LeBlanc \& Wilson will not propagate to the stellar
surface as a jet.  The calculation of Khokhlov et al. (1999) 
shows that this is not necessarily correct, at least for a helium core.
Meier et al. based their analysis on stellar evolution calculations of
the day, but they adopted a stellar core with central density of
about $10^{10}$~\gcm3 giving a binding energy of about $10^{52}$ ergs.
This exaggerates the binding energy of the initial core by about a
factor of 10 compared to modern calculations and gives an incorrectly
small value of a key parameter of Meier et al., the ratio of the
binding energy of the newly-formed neutron star to that of the
initial core.  Meier et al. also did not consider the possibility of
an $\alpha-\Omega$ dynamo that could lead to exponential field growth
(Duncan \& Thompson 1992).  Symbalisty found that both substantial
rotation and magnetic field were necessary to affect the explosion
and that the presence of the magnetic field and associated losses
allowed deeper collapse.  The whole question of the initiation of
MHD jets in association with neutron star formation needs to be
considered anew.  Here we consider some basic properties of
jet propagation.                                         

The jet would be stopped by the envelope when the jet is unable to
provide the power to move the envelope material at a speed 
comparable to the jet velocity.  This can be expressed as:
\begin{equation}
\label{stopjet}
L_{jet}\simeq R^2\Delta\Omega\times\rho_{env}v_{jet}^2\times v_{jet},
\end{equation}
where $\Delta\Omega$ is the solid angle of the jet.
The jet would then be stopped in a length
\begin{equation}
\label{stoplength}
R_{st}\simeq \left[\frac{L_{jet}}
{\Delta\Omega\rho_{env}v_{jet}^3}\right]^{1/2}.
\end{equation}
In order to penetrate the star, the energy injected into the envelope at 
$R\lta R_{st}$ 
should be enough to unbind the region of the outer stellar mantle
impacted by the jet.  
The amount of stellar material impacted by the jet is
\begin{equation}
\label{mpush}
\Delta M_{env} = M_{env}\frac{\Delta\Omega}{4\pi}
\simeq8\times10^{-3}M_{env}\left(\frac{\Delta \Omega}{0.1}\right).
\end{equation}
Unbinding this mass requires
\begin{equation}
\label{unbind1}  
L_{jet}\gta\frac{GM_{env}M_{NS}\Delta\Omega}{4\pi R_{env}\Delta t},
\end{equation} 
where $\Delta t$ is the duration of injection of the jet.  
Equation (\ref{unbind1}) can be expressed as:
\begin{equation}
\label{unbind2}
L_{jet}\gta3\times10^{47} {\rm erg~s^{-1}}
\left(\frac{M_{env}}{1M_\odot}\right)\left(\frac{M_{NS}}
{1.5 M_\odot}\right) \left(\frac{R_{env}}{10^{5}~{\rm km}}\right)^{-1}
\left(\frac{\Delta\Omega}{0.1}\right)\left(\frac{\Delta
t}{1 s}\right)^{-1}.
\end{equation} 
A jet of $\simeq10^{51}$~\erg-s thus gives ample power
to unbind a portion of the envelope occupying about 0.1 sterradian.


The dynamics of the jet and impacted envelope material will depend
on whether the mass of the jet is greater than or less
than the mass of the displaced stellar envelope.
The speed of the jet will be
\begin{equation}
\label{vjet}
v_{jet}\simeq2 c \left(\frac{2L_{jet}\Delta t}{M_{jet}c^2}\right)^{1/2}.
\end{equation}      
This implies a relativistic velocity for $L\Delta t\simeq10^{51}$ ergs
and $M_{jet}\lta0.01\m$.  If a comparable amount of
energy is put into the displaced envelope material,
the impacted envelope mass, $\Delta M_{env}$, would expand at a speed 
\begin{equation}
\label{veject1}
v_{ej}\simeq\left(\frac{2L_{jet}\Delta t}{M_{env}}\
\frac{4\pi}{\Delta\Omega}\right)^{1/2}.
\end{equation}
If this matter were displaced radially out of the star, it 
could also acquire relativistic speeds for  $L\Delta t\simeq10^{51}$ ergs 
and $\Delta\Omega\lta0.1$.  In practice, the displaced material will tend to
be accelerated sideways by shocks induced by the passage of the jet
and the energy will be shared by the whole envelope at roughly the
sound speed in the envelope.  

The propagation of the jet requires more study, but the calculation
of Khokhlov et al. (1999) gives some qualitative insight into
the behavior to be expected.  As the jet propagates, a bow shock
runs ahead of it.  The speed of the bow shock is less than that
of the matter inflow into the jet, a ratio of about 0.5 to 0.7
for the particular
case explored by Khokhlov et al., but that ratio will depend on 
the speed and density of the jet, its opening angle, and the 
structure of the star through which the jet propagates.

The bow shock of the jet will both heat material and 
cause it to expand sideways.
The opening half angle of the jet will then be approximately
\begin{equation}
\label{jetangle}
\theta \simeq \frac{v_{env}}{v_{bow}}\simeq 0.1~{\rm rad}
\frac{v_{env,8}}{v_{bow,9}}\simeq 5^o\frac{v_{env,8}}{v_{bow,9}}.
\end{equation} 
For $v_{env}\simeq2\times10^8$ cm s$^{-1}$ and $v_{bow}\simeq2\times10^9$
cm s$^{-1}$, $\Delta\theta \simeq 5^o$  
or $\Delta\Omega/4\pi\simeq 0.004$. 
The actual dynamics of the jet will depend on nested cocoon-like
shocks from the bow shock and subsequent expansion, as illustrated
by Khokhlov et al. (1999).

\section{The Neutron Star Phase}

Concurrent with the propagation of any MHD jet and magnetic 
torque exerted by the proto-neutron star, the neutron star will
contract and new physics can come into play.
We assume that the cooling and contraction leads to a neutron 
star rotating at a period of about 1 millisecond.  
This implies a much smaller radius for the light cylinder,
\begin{equation}
\label{lcylinderNS}  
R_{LC} = \frac{c}{\Omega_{NS}}\simeq50~{\rm km}
\left(\frac{P_{NS}}{{\rm 1~ms}}\right).
\end{equation}
This is only a little bigger than the neutron star radius.
If the field remains about the same as the proto-neutron star,
$B\simeq10^{14}$ G, the Alfv\'en radius will be comparable to $R_{LC}$, but
if the field amplifies during the contraction, as is plausible, the
Alfv\'en radius will expand to become substantially greater than 
$R_{LC}$.  Under this circumstance, a stationary dipole configuration 
at and beyond the Alfv\'en radius can no longer be maintained.

During the contraction phase, the Rossby number decreases in proportion
to the rotational period. Assuming the neutron star convective time
scale remains at about 1 ms, the condition $R_0\lta1$ will be reached
if the neutron star ends up with a period of about 1 ms.
This means that an $\alpha-\Omega$-type dynamo could be initiated
that would build a strong magnetic field up to $\sim10^{17}-10^{18}$ G
(Duncan \& Thompson 1992; Thompson \& Duncan 1993; equipartition 
with the rotation would give $\sim10^{18}$ G).  In this extreme
circumstance, the surface field could be
$\sim10^{15}-10^{16}$ G for the dipole configuration near the 
neutron star surface (in the absence of the light cylinder).  
This phase of field growth
occurs exponentially in contrast to the linear amplification
associated with the relatively slowly-rotating initial proto-neutron star.
If the dipole field grows to $B_{NS}\simeq10^{16}$ G as the radius shrinks
to $R_{NS}\simeq10^{6}$ cm, then, all else being equal, the
Alfv\'en radius will be $\simeq140$ km, substantially larger than
$R_{LC}$, as just noted.  If the ram pressure has declined,
$R_{A}$ will be even larger.
 
The rotational energy after the contraction is
\begin{equation}
\label{ErotNS}  
E_{rot,NS}\simeq\frac{1}{2}I_{NS}\Omega_{NS}^2
\simeq6\times10^{52}~{\rm erg}
\left(\frac{M}{1.5 {\m}}\right)
\left(\frac{\Omega_{NS}}{10^4~{\rm s^{-1}}}\right)^2
\left(\frac{R_{NS}}{10^6~{\rm cm}}\right)^2.
\end{equation}
Setting aside the baryon-loading problem for the moment, this
energy is comparable to the largest energy associated with 
any \grb, $\sim 3\times10^{54}\Delta\Omega/4\pi$ erg for
GRB~990123 (Kulkarni, et al. 1999) for a degree of collimation,
$\Delta\Omega/4\pi\lta10^{-2}$.  This
degree of collimation has been deduced for some afterglows,
in particular for GRB~990123 itself,
and as noted in the previous section, 
is about the order expected for a matter-dominated jet (see also 
Khokhlov et al. 1999). The degree of
collimation of the subsequent flow of electromagnetic radiation
is unclear.  We return to that topic below.

The rotational energy of the contracted neutron star
is radiated away in the form of intense waves of
frequency $\Omega_{NS}$ generated at the speed of
light circle.  To generate traditional propagating 
electromagnetic waves as opposed to MHD waves,
the displacement current must exceed the plasma current.  This
criterion can be written roughly as
\begin{equation}
\label{displace}
\rho \lta \frac{B}{e c  N_0 P_{NS}} 
\simeq 10^{-6}~{\rm g~cm^{-3}}\left(\frac{B_{NS}}{10^{16}~{\rm G}}\right)
\left(\frac{P_{NS}}{1~{\rm ms}}\right)^{-1}, 
\end{equation}
where e is the electron charge and $N_0$ is Avogodro's number.
The density within the star vastly exceeds this limit, so
the energy will be generated as MHD waves.
These waves will have extreme properties that require futher
study.  In particular, the Alfv\'en speed can be estimated as 
\begin{equation}
\label{alfvenspeed}
v_A = \left(\frac{B^2}{8\pi\rho}\right)^{1/2} \simeq
2\times 10^{11}~{\rm cm~s^{-1}}\left(\frac{B_{NS}}{10^{16}~{\rm G}}\right) 
\left(\frac{\rho}{10^{8}~{\rm g~cm^{-3}}}\right)^{-1/2}. 
\end{equation}      
If the surface field is of order $10^{16}~{\rm G}$, then the
field at the speed of light circle may be about $10^{15}~{\rm G}$.
Even for weaker fields, the density near the speed of light
circle will decline in dynamic reaction to the large deposition
of pulsar energy.  Equation (\ref{alfvenspeed}) thus serves to
argue that this simple Newtonian expression for the speed of
MHD waves is likely to be invalid since it is neither relativistic
nor applicable in the strong field limit where the waves are
not a first order perturbation to the plasma.  Instead
one must treat the generation and propagation of high amplitude,
ultrarelativistic MHD waves (UMHDW), a task beyond the scope of this paper.
Rather, we will sketch the basic energetics and the potential
behavior of these waves.  

If the density gets sufficiently low as the UMHDW propagate
or because of dynamical reaction to energy deposition that
creates a cavity causing the density to drop,
then the plasma current can be exceeded by the displacement
current and the waves can be described as Large Amplitude
Electromagnetic Waves (LAEMW; Usov 1994; Blackman \& Yi 1998).
The properties of these waves are also not completely understood.
Any LAEMW cannot propagate through a plasma if the plasma frequency
\begin{equation}
w_p~=~\left(\frac{4\pi n_e e^2}{m_e}\right)^{1/2}
\simeq 9\times10^3n_e^{1/2}~{\rm Hz},
\end{equation}
exceeds the LAEMW frequency $\Omega_{NS}$.  The corresponding
condition on the electron density is:  
\begin{equation}
\label{edensity}  
n_e \gta 1~{\rm cm^{-3}}\left(\frac{\Omega_{NS}}{10^4~{\rm s^{-1}}}\right)^2.
\end{equation}
For densities exceeding this value, the LAEMW would have a very
small skin depth and cannot propagate in the plasma.  
They would effectively be reflected by the plasma.  The density
exceeds this critical value under any stellar conditions.

In the discussion below we neglect the production of traditional 
Alfv\'en waves as an energy loss mechanism.  Alfv\'en waves occur when the
magnetic field is relatively weak and the waves are a propagating
perturbation.  Alfv\'en waves will thus be generated beyond the  
magnetosphere where the magnetic field no longer dominates the 
particle dynamics.  In the proto-neutron star case, Alfv\'en waves 
could be generated in the region beyond the magnetosphere
and within the extended light cylinder.  The torque acting at
$R_A$ probably dominates those losses, but this is worth more detailed
consideration.  Once the neutron star contracts and the
light cylinder passes within the magnetosphere, 
where Alfv\'en waves are not a relevant concept, 
the dominant losses will occur at the speed of light circle 
by generation of UMHDW.  Alfv\'en waves might be generated
up the rotation axis within the light cylinder but beyond
the Alfv\'en radius.  The global behavior of the open magnetic
field lines that cross the light cylinder
in the complex environment we are discussing is
unclear.  The dynamics of these open field lines 
again might lead to the production of Alfv\'en
waves and associated losses, but it seems likely that in all
these possibilities, Alfv\'en waves remain a secondary process.

The luminosity in UMHDW is estimated to be
\begin{equation}
\label{LNS1}
L_{MHD}\simeq4\pi R_{LC}^2\times\frac{c}{4\pi}|\vec E\times\vec B|
\simeq \frac{\mu_{NS}^2}{R_{LC}^4}c
\simeq\frac{R_{NS}^6B_{NS}^2 \Omega_{NS}^4}{c^3},
\end{equation}
assuming the UMHDW to be generated at $R_{LC}$ and the magnetic moment
of the neutron star to be $\mu_{NS}=B_{NS}R_{NS}^3$. For the conditions
of the contracted neutron star which has initiated an $\alpha-\Omega$
dynamo, we expect
\begin{equation}
\label{LNS2}  
L_{MHD}\simeq4\times10^{52}~{\rm erg~s^{-1}}
\left(\frac{R_{NS}}{10~{\rm km}}\right)^6
\left(\frac{B_{NS}}{10^{16}~{\rm G}}
\right)^2\left(\frac{\Omega_{NS}}{10^4~{\rm s^{-1}}}\right)^4,
\end{equation}
which will last for a duration of
\begin{equation}
\label{temission}  
t_{MHD}\simeq\frac{E_{rot,NS}}{L_{MHD}}\simeq 2~{\rm s}
\left(\frac{M}{1.5~{\m}}\right) 
\left(\frac{R_{NS}}{10~{\rm km}}\right)^{-4}
\left(\frac{B_{NS}}{10^{16}~{\rm G}}\right)^{-2}
\left(\frac{\Omega_{NS}}{10^4~{\rm s^{-1}}}\right)^{-2}.
\end{equation}

The UMHDW are likely to initially be bottled up near the site of their 
production near $R_{LC}$ since the surrounding plasma can respond
only at the subrelativistic sound speed.  
The UHMDW would push the plasma aside 
and seek the easiest way out.  An obvious possibility is that they
will ``burn" a channel along the rotation axis, following the path of
the previous MHD jet.
We assume that the most intense production of UMHDW 
occurs after the completion of the contraction of the
neutron star, and hence that it is delayed by the cooling 
time $\simeq$ 5 to 10 s, about the time necessary for the initial matter 
jet to propagate through the stellar core. 
A mass of about $\Delta M\simeq
M_{env}\frac{\Delta\Omega}{4\pi}\simeq10^{-2}\ M_{env}$  
will be pushed aside by the initial matter jet.
Whether the region impacted by the matter jet will be rarefied depends
on the mass of the jet.  If the jet is rather massive, as in
the case of the calculation of Khokhlov et al. (1999), then 
the jet remains denser than the stellar environment as it
propagates into the envelope. In this case, the jet acts
as a ``plug" until it disperses after several dynamical times. 
The UHMDW generated by the pulsar will tend
to propagate through the lowest density regions.  If there is
a density minimum along the rotation axis, there will be 
a tendency for energy to flow in that direction.  
The dynamics are likely to be complicated and we will
just sketch the possibilities.

The UMHDW should be rapidly isotropized in the co-moving frame, 
so they can be considered as a relativistic gas while
they are trapped within the stellar core.  It is not so clear
that they are thermalized since their characteristic
wavelength
\begin{equation}
\label{wavelength}
\lambda_{MHD}\simeq cP_{NS}\simeq 2\pi R_{LC} \simeq 300~{\rm km},
\end{equation}
is comparable to or larger than the scale of the region in which
they are generated.  This is an important issue since
if the energy is thermalized then the effective temperature
will be,
\begin{equation}
\label{radtemp}
T_{MHD}\simeq\left(\frac{E_{MHD}}{\frac{4\pi}{3}aR^3}\right)^{1/4}
\simeq2\times10^{10}~{\rm K}~E_{52}^{1/4}R_3^{-3/4},
\end{equation}
where $R_3$ is the radius in units of $10^3$ km,
so that much of the energy would go into the formation of
pairs which might be dissipated by adiabatic expansion
and difficult to recover.  
We will ignore this possibility
for the moment, assuming that the energy cannot thermalize.

If the UHMDW act like a relativistic gas, then they will
exert a pressure of
\begin{equation}
\label{radpress}
P_{MHD}\simeq\frac{1}{3}
\left(\frac{E_{MHD}}{\frac{4\pi}{3}aR^3}\right)
\simeq8\times10^{26}~{\rm erg~cm^{-3}}~E_{52}R_3^{-3}.
\end{equation}    
This may be contrasted to the pressure required to ensure 
hydrostatic equilibrium in the gravity of the neutron star which
we can represent crudely as,
\begin{equation}
\label{PHSE}
P_{HSE}\simeq\frac{GM_{NS}^2}{R^4}
\simeq 1\times10^{28}~{\rm erg~cm^{-3}}\left(\frac{M_{NS}}{1.5 \m}\right)^2
R_3^{-4}.
\end{equation}          
By this measure, the pressure of the radiation will be less than
required to produce HSE for small radii, and will exceed the
pressure corresponding to HSE at a radius of 
\begin{equation}
\label{RHSE}
R_{HSE}\simeq1\times10^{4}~{\rm km}\left(\frac{M_{NS}}{1.5 \m}\right)^2
E_{52}^{-1}.
\end{equation}   
The regime where the UMHDW begin to dominate will, of course, 
depend on the density profile in the collapsing matter and 
the whole environment is, in any case, dynamic so that considerations
of HSE give only a partial perspective.

The production of the UMHDW near
the light cylinder may lead to strong Rayleigh-Taylor instability
as the relativistic fluid pushes on the surrounding gas.  The 
subsequent behavior will depend on the rate at which the UMHDW ``gas"
expands quasi-spherically in Rayleigh-Taylor fingers compared to
the rate at which the waves will selectively propagate upward
along the rotation axis.
It seems plausible that a substantial majority of the
energy will flow up the axis, the path of least resistance.
Another possibility is that the energy density becomes so high
compared to the surrounding matter, near the neutron star or
further out in the mantle (equation \ref{RHSE}), 
that the stellar matter becomes an
insubstantial barrier to the nearly free expansion of the UMHDW
bubble.  Since the total energy in the UMHDW could be as high as
$10^{52}$ erg, this is a real possibility.  Whether the UMHDW
remain collimated may thus depend sensitively on the timescale on
which they are generated and hence the instantaneous energy density
throughout the star.  These possibilities clearly need to be investigated 
with appropriate numerical calculations.  

The result of the generation of the UMHDW could be rather different
if they are thermalized.  In this case, the UMHDW would be
replaced with copious pairs (cf. equation \ref{radtemp}).  The
pair gas would also act like a relativistic gas with the same
issues of dynamics just discussed.  In addition, there would
be the complications of annihilation at boundaries with normal
matter and the possibility of strong adiabatic losses after
a pair jet broke out of the star and underwent free expansion.
Such loss of pair thermal energy to rest mass and kinetic
energy might be at least partially recovered if the pairs
interacted with a surrounding wind of normal matter (see \S 8).



\section{Evolution of Energy Emission and Gamma-Ray Bursts}

There are two stages when \grs\ might be generated.  The
first is when the bow shock that proceeds the initial mass-dominated jet
impacts on the stellar photosphere.  The second phase is when
a collimated flow of UMHDW erupts from the surface at about
the same time. We consider those in turn.

\subsection{Bow Shock Gamma-Ray Emission}

Some \gr\ emission may be generated by the phase of shock breakout
as the shock associated with the initial matter-dominated jet 
runs down the stellar 
density gradient and breaks through the photosphere.  An associated
mechanism is for the accelerated matter to reach relativistic speeds
and then collide with some external matter.

As the bow shock that proceeds the jet runs down the exponential stellar 
density gradient in the photosphere it will accelerate, in turn heating 
and accelerating the matter, a mechanism to produce \grs\ first described
by Colgate (1974, 1975; see Matzner \& McKee 1999 for an updated
discussion of analytic models of shock breakout).  
The question of how this matter expands 
and radiates and perhaps collides with an external environment
is beyond the scope of this paper, but worth more detailed study.
Here we remark on the basic properties at the time of shock passage.

Energy is deposited in the atmosphere by the bow shock.  The
bow shock, in turn, derives its energy from the jet. The power
carried by the jet is $\simeq 1/2\rho_{jet} v_{jet}^3\Delta A$,
where $\Delta A$ is the fractional area
of the star that is impacted by the bow shock.  This power
will be delivered to the photosphere in a time $\simeq l_{phot}/v_{jet}$,
where $l_{phot}$ is the depth of the photosphere.
The energy deposited in the photosphere during the bow shock
break out phase is thus:  
\begin{equation}
\label{atmenergy}    
E \simeq \frac{1}{2}\rho_{jet} v_{jet}^2 l_{phot} \Delta A 
\simeq 5\times10^{45}~{\rm erg~} 
\rho_{jet}v_{jet,10}^2 l_{phot,7}\Delta A_{18},   
\end{equation}
where $v_{jet,10}$ is the velocity of the bow shock in units of
$10^{10}$ \cms,  $l_{phot,7}$ is in units of $10^7$ cm and $\Delta A_{18}$ is 
in units of $10^{18}$ cm$^2$.  For the calculation of Khokhlov
et al. (1999) the jet density is about $\rho_{jet} \simeq 10^3$ \gcm3,
and the velocity steepens to about $90,000$ \kms as the jet
approaches the outer density gradient of the helium core at which point
the resolution of the density profile degrades.
For $\rho_{jet} \simeq 10^3$ \gcm3, $v_{jet,10}\simeq 1$,
$l_{phot,7} \simeq 1$, and $\Delta A_{18} \simeq 1$, 
the energy would be $\simeq 5\times 10^{48}$
ergs, comparable to the \gr\ energy in the burst of SN~1998bw/GRB~980425.

The corresponding temperature, making the crude, and not 
necessarily correct, assumption of thermalization to a 
radiation-dominated gas, is
\begin{equation}
\label{bowtemp}    
T \simeq \left(\frac{\rho_{jet}v_{jet}^2}{2a}\right)^{1/4} 
\simeq 10^8~{\rm K}~\rho_{bow}^{1/4} v_{jet,10}^{1/2},
\end{equation}
For $\rho_{jet} \simeq 10^3$ \gcm3, $v_{jet,10}\simeq 1$,    
equation (\ref{bowtemp}) gives $T\simeq 10^{10}$ K with the possible
emission of \grs.    

This hot, accelerated matter would then expand and radiate.  
Whether it could account for the \grb\ in SN~1998bw/GRB~980425
will require more careful consideration.  We note that if this
energy accumulates in the density gradient of the photosphere 
the matter will be optically thin, 
so not susceptible to adiabatic losses while it undergoes free
expansion. The photosphere is already optically
thin by definition and the opacity will be decreased by Klein-Nishina
corrections. On the other hand, the jet is denser and hence
more opaque than the photosphere so it is important to 
know where the energy represented by equation (\ref{atmenergy})
resides. Although the conditions could vary widely depending
on the nature and propagation of the jet, for the calculation of
Khokhlov et al. (1999), the bow shock material is matter ablated from
the jet and the characteristic density is about the same as the jet,
$\rho_{bow} \simeq 10^3$ \gcm3.  

The emission properties of the jet/bow shock/stellar atmosphere region 
as the jet impacts the atmosphere thus require more careful study.
Nevertheless, considerable hard radiation could be emitted 
before the phase of homologous
expansion is reached.  The time scale for this energy to
be radiated is also of importance.  The shock break out time,
$\simeq 10^{-3}~l_{phot,7}/v_{jet,10}$ s, is far too short to correspond
to an observed \grb, specifically that in GRB~980425, but
the controlling time scale will be the radiative timescale.

Another possible means of producing \grs\ from the matter
jet is to accelerate
the matter in the photosphere to relativistic speeds and
for it then to subsequently collide with a surrounding medium,
perhaps a stellar wind.  The mass fraction that can be accelerated
in this way for an
explosion with an impulsive energy input that drives a single
shock through the outer stellar layers has been 
evaluated for CO and He stellar cores by Woosely, Eastman \& Schmidt
(1998).  For their models, they establish a relationship
\begin{equation}
\label{Q}  
Q = \Gamma \beta \left(\rho r^3 \right)^{0.2} 
\simeq 2.5\times10^6~\Gamma M_{ex,32}^{0.15},
\end{equation}
where the parameter Q is independent of Lagrangian mass
and radius in the ejecta, $M_{ex,32}$ is the mass external to
the layer with Lorentz factor $\Gamma$ in units of $10^{32}$ gm, 
and we have taken
$\beta \simeq 1$.  Woosely et al. find that $Q \simeq 3\times 10^5$.
At the risk of overinterpreting their results, we note that
for two otherwise identical models that differ only in the
explosion energy input (models CO6A and CO6C), Q 
scales with the explosion energy, $E_{ex}$, approximately as $E_{ex}^{1/2}$.  
In the following we have adopted $Q \simeq 10^5 E_{ex,51}^{1/2}$,
where $E_{ex,51}$ is in units of  $10^{51}$ erg.

With this scaling, we find the kinetic energy in the homologously
expanding material (e.g. after the immediate shock heating phase
discussed above) to be  
\begin{equation}
\label{KEaccel}   
KE = \Gamma M_{ex} c^2 \simeq 1.6\times10^{44}~{\rm erg}~
\Gamma^{-5.66}E_{ex,51}^{2.83}.
\end{equation}
This expression is for a spherical explosion.  It is clear
that the energy in matter accelerated to relativistic speeds is
insufficient to account even for the weak \grb\ imputed to GRB~990425,
as concluded by Woosely, Eastman \& Schmidt (1998).  

In the present
context, it is relevant to estimate the difference if the
shock did not propagate spherically, but were collimated.  
In this case, we are interested in the kinetic energy in
a jet with solid angle $\Delta \Omega/4\pi$ or 
$KE_{\Delta \Omega} = KE\Delta \Omega/4\pi$.  The relevant 
input energy is the energy in the jet, 
$E_{jet} = (\Delta \Omega/4\pi) E_{ex}$, where $E_{ex}$ is now
to be thought of as the equivalent isotropic energy of the collimated
jet.  Making this substitution in equation (\ref{KEaccel}), we find:
\begin{equation}
\label{KEomega}
KE_{\Delta \Omega} \simeq 7.2\times10^{47}~{\rm erg}~
\Gamma^{-5.66}f_{coll-2}^{-1.83} E_{jet,51}^{2.83},
\end{equation}                               
where $f_{coll-2} = \Delta \Omega/4\pi$ in units of $10^{-2}$.
Equation (\ref{KEomega}) shows that even with substantial collimation,
the amount of energy put into
mildly relativistic matter ($\Gamma \gta$ a few) 
by this mechanism is still insufficient to account for the
putative \grb\ in SN~1998bw for an energy typical of the
matter jet we have discussed here, $E_{jet,51} \simeq 1$.  There
might be sufficient energy for $E_{jet,51} \simeq 10$, but
even this energy would be insufficient to boost enough stellar
matter to $\Gamma \simeq 10$ to account for SN~1998bw.  

From this analysis, we conclude that, even if it is strongly
collimated, the physical process of running a single impulsively-induced
shock (e.g. Colgate 1975; Matzner \& McKee 1999) 
down the photospheric density gradient of a compact 
stellar core, accelerating that matter to relativistic speeds,
and slamming it into a surrounding medium 
is unlikely to generate a substantial \grb\ of
the kind associated with SN~1998bw.  To produce a cosmic \grb\
with an energy of $10^{52}$ erg (isotropic equivalent of $10^{54}$ erg)
and $\Gamma \gta 100$ is out of the question.  If jets from 
stellar collapse to produce either neutron stars or black holes
are to generate \grbs\ in SN~1998bw, never mind the high redshift
events, then the physics of the \gr\ production must be associated
with the initial shock breakout phase (as discussed above) or
by the prolonged emission phase of the jet passing through
and beyond the photosphere. 

\subsection{The Contraction/UMHDW Phase}

We now consider the time evolution of the contracting neutron star
and the associated energy emission mechanisms implied by this 
scenario.

When $R_{LC}\gg R_A$, the magnetic torque is the dominant energy/momentum
transport mechanism (\S 4.2).  When $R_{LC}\ll R_A$, Poynting
flux/UMHDW transports the energy/momentum.  
In our scenario, the proto-neutron star
continues spinning up due to contraction even though
it transfers angular momentum through magnetic torque.  Initially,
when $R_{LC}\gg R_A$, the rate of loss of energy to 
electromagnetic Poynting flux
would be (cf. equation \ref{LNS2})
\begin{equation}
\label{EMpower}  
L_{EM}\simeq~2\times10^{46}~{\rm erg~s^{-1}}
\left(\frac{R_{PNS}}{50~{\rm km}}\right)^6
\left(\frac{B_{PNS}}{10^{14}~{\rm G}}\right)^2
\left(\frac{\Omega_{PNS}}{250~{\rm s^{-1}}}\right)^4.
\end{equation}
This will be small compared to the energy deposited by the
torque of the spinning neutron star acting at the Alfv\'en
radius (equation \ref{torqueenergy}),
\begin{equation}
\label{torquepower}
L_{tor}\simeq1\times10^{49}~{\rm erg~s^{-1}}
\left(\frac{\Omega_{PNS}}{250~{\rm s^{-1}}}\right)
\left(\frac{R_{PNS}}{50~{\rm km}}\right)^3
\left(\frac{B_{PNS}}{10^{14}~{\rm G}}\right)
\left(\frac{\rho}{10^8~{\rm g~cm^{-3}}}\right)^{1/2}
\left(\frac{v}{10^8~{\rm cm~s^{-1}}}\right).
\end{equation}                                                      

After contraction
when the condition $R_{LC}\lta R_A$ is reached, we have:
\begin{equation}
\label{alfvenNSratio}
\frac{R_A}{R_{NS}}\simeq~14\left(\frac{B_{NS}}{10^{16} 
~{\rm G}}\right)^{1/3}
\left(\frac{\rho}{10^8~{\rm g~cm^{-3}}}\right)^{-1/6}
\left(\frac{v}{10^8~{\rm cm~s^{-1}}}\right)^{1/3},
\end{equation}
and 
\begin{equation}
\label{lcylinderNS2}  
R_{LC}\simeq~30~{\rm km}\left(\frac{\Omega_{NS}}
{10^{4}~{\rm s^{-1}}}\right)^{-1},
\end{equation}
and hence
\begin{equation}
\label{lcylinderalfven}  
\frac{R_{LC}}{R_A}\simeq0.2
\left(\frac{R_{NS}}{10~{\rm km}}\right)^{-1}
\left(\frac{B_{NS}}{10^{16}~{\rm G}}\right)^{-1/3}
\left(\frac{\rho}{10^8~{\rm g~cm^{-3}}}\right)^{1/6}
\left(\frac{v}{10^8~{\rm cm~s^{-1}}}\right)^{1/3}
\left(\frac{\Omega_{NS}}{10^4~{\rm s^{-1}}}\right)^{-1},
\end{equation}
where $\rho$ and $v$ are to be evaluated at the Alfv\'en radius.
Since energy injection during the phase when $R_{LC}>R_A$ 
will drive the density down in the vicinity of the
magnetopause and the standing shock, and 
$B_{NS}$ is likely to increase from $10^{14}$~G to
$10^{16}$~G continuously (but suddenly) during the contraction
by the linear amplification mechanism and dynamo, 
the epoch during contraction 
when $R_{LC}\simeq R_A$ will occur is difficult to
estimate; however, since essentially all the energy emitted
will be used to expel the envelope (i.e. either by torque or UMHDW)
a more significant
explosion/expansion is expected when $R_{LC}$ $\lta$ $R_A$.  At this epoch
we expect that if the torque were still active the rate of
loss of rotational energy due to this mechansim would be
\begin{equation}
\label{torquepowerNS}
L_{tor}\simeq3\times10^{49}~{\rm erg~s^{-1}}
\left(\frac{\Omega_{NS}}{10^4~{\rm s^{-1}}}\right)
\left(\frac{R_{NS}}{10~{\rm km}}\right)^3
\left(\frac{B_{NS}}{10^{16}~{\rm G}}\right)
\left(\frac{\rho}{10^8~{\rm g~cm^{-3}}}\right)^{1/2}
\left(\frac{v}{10^8~{\rm cm~s^{-1}}}\right).
\end{equation}
This power would in any case be overwhelmed by the loss of 
energy to the UMHDW with the enhanced rotation and magnetic field,
\begin{equation}
L_{MHD}\simeq~4\times10^{52}~{\rm erg~s^{-1}}
\left(\frac{R_{NS}}{10~{\rm km}}\right)^6
\left(\frac{B_{NS}}{10^{16}~{\rm G}}\right)^2
\left(\frac{\Omega_{NS}}{10^4~{\rm s^{-1}}}\right)^4.
\end{equation}
This power output will last for a few seconds (equation \ref{temission}). 
The injected energy is then 
$\gta$ 10$^{52}$ erg, which is enough to drive a significant shock.  
In this scenario, both a \grb\ and an asymmetric supernova could 
be produced during the ``magnetar" phase.

The UMHDW will likely
flow up the rotation axis starting with a cross-sectional area
of $\simeq R_{LC}^2$ and rising as an intense photon ``bubble." 
Assuming that this bubble pushes matter aside at roughly the
local sound speed, the opening angle will be: 
\begin{equation}
\label{angleLAEW}
\theta_{MHD}\simeq\frac{c_s}{c}\simeq 3\times10^{-3}~{\rm rad}
\simeq 0.2^o,
\end{equation}   
where $c_s$ is the sound speed in the helium core, 
$\simeq 10^8$ \cms.  This corresponds to a solid angle of 
$\Delta\Omega/4\pi\simeq2\times10^{-6}$.
The UMHDW may then propagate up the axis in a radiation-dominated
jet with smaller cross section than the original matter-dominated jet.  
If this is the case,
the channel carved by the UMHDW may remain substantially 
smaller than the characteristic wavelength of the UMHDW until
the waves break out of the surface of the star.

To avoid the baryon loading problem and to generate 
a \grb\ with large Lorentz factor, the hole through which the
bulk of the UMHDW emerge should contain less than  
\begin{equation}
\label{baryonload1} 
\Delta M~\lta~\frac{10^{52}~{\rm ergs}}{\Gamma^2 c^2},
\end{equation}
(Rees \& M\'esz\'aros 1992).
For $\Gamma\gta100$, the requirement is 
$\Delta M\lta10^{27}~{\rm g}\simeq5\times10^{-7}$~\m. 
It is difficult to estimate whether the UMHDW jet will
entrain such a small amount of matter.  There are
two qualitative possibilites for the propagation of
the UMHDW jet.  There may be a density minimum along the
axis of the previous matter jet as suggested by the calculations
of Khokhlov et al (1999).  If the UMHDW propagate through such a low
density axial channel, they will naturally emerge somewhat
more collimated than the original jet.  On the other hand,
if the jet is denser than the surrounding stellar 
matter and the jet acts like a plug on the axis, the UMHDW
might propagate in a cylindrical blanket around the plug.
They would then emerge with an annular cross section.
This would result in yet another complication for 
predicting the observational aspects of such an event.

If the matter jet leaves behind a
region of very low baryon density, or the UHMDW emerge
beyond the end of the more slowly propagating matter jet,
then the propagating 
UMHDW could enter a low density region in which the density
of the environment is lower then the critical Goldreich-Julian density
(Goldreich \& Julian 1969; Shapiro \& Teukolsky 1983) of
\begin{equation}
\label{GJdensity}
\rho_{GJ}\simeq 10^{-6}~{\rm g~cm^{-3}}
\left(\frac{B(R)}{10^{16}~{\rm G}}\right)
\left(\frac{P}{1~{\rm ms}}\right)^{-1},
\end{equation}                                   
where the flux-freezing and force free conditions are broken
(see equation \ref{displace}).  At this point, the transition
from UMHDW to LAEMW can be made.  Reconnection
of magnetic fields and rapid acceleration of particles in a pair plasma
follows (e.g. Asseo, Kennel \& Pellat 1978; Usov 1992; Michel 1984; 
Thompson 1994; Blackman, Yi \& Field 1995). 
The pair plasma could in principle be accelerated to 
a high bulk Lorentz 
factor up to $\simeq 10^5$, although the exact 
value of the maximum bulk Lorentz
factor could be significantly lower than this due to baryon loading and
other complicated factors reducing acceleration efficiency 
(cf. Usov 1992, 1994).



The rest-frame \gr\ luminosity would be 
determined by the electromagnetic power, i.e.
\begin{equation}
\label{Lobs1}
L_{obs}\simeq f_{\gamma}L_{MHD}\simeq4\times10^{51}~{\rm erg~s^{-1}}  
\left(f_{\gamma}\over 0.1\right)
\left(\Omega_{NS}\over 10^4~{\rm s^{-1}}\right)^4
\left(R_{NS} \over 10~{\rm km}\right)^6
\left(B_{NS}\over 10^{16}~{\rm G}\right)^2,
\end{equation}
where $f_{\gamma}$ is the \gr\ emission efficiency factor.
During this phase, the pulsar rotation energy is
used to power the \grb\ while the spin of the pulsar slows down as
\begin{eqnarray}
\label{spindown}
\Omega_{NS}&=&10^4~{\rm  s^{-1}}\left(\Omega_{NS,i}\over 
10^4~{\rm s^{-1}}\right)\\
&&\mbox{}\times\left[1+6\times10^{-2}
\left(M_{NS} \over 1.5 {\m}\right)^{-1}
\left(\Omega_{NS,i}\over 10^4~{\rm s^{-1}}\right)^2
\left(R_{NS} \over 10~{\rm km}\right)^4
\left(B_{NS}\over 10^{16}~{\rm G}\right)^2
\left(t-t_{em,i} \over 1~{\rm s}\right)\right]^{-1/2},\nonumber
\end{eqnarray}
where $\Omega_{NS,i}$ is the initial spin 
frequency of the neutron star and
$t_{em,i}$ is the initial time of the spin-down phase driven by the
electromagnetic dipole radiation emission. The time evolution of the
\gr\ luminosity is then given by
\begin{eqnarray}
\label{Lobs2}
L_{obs}&\simeq&4\times10^{51}~{\rm erg~s^{-1}}   
\left(f_{\gamma}\over 0.1\right)
\left(\Omega_{NS,i}\over 10^4~{\rm s^{-1}}\right)^4
\left(R_{NS} \over 10~{\rm km}\right)^6
\left(B_{NS}\over 10^{16}~{\rm G}\right)^2 \\
&&\mbox{}\times\left[1+6\times10^{-2}
\left(M_{NS} \over 1.5 {\m}\right)^{-1}   
\left(\Omega_{NS,i}\over 10^4~{\rm s^{-1}}\right)^2
\left(R_{NS} \over 10~{\rm km}\right)^4
\left(B_{NS}\over 10^{16}~{\rm G}\right)^2
\left(t-t_{em,i}\over 1~{\rm s}\right)\right]^{-2}.\nonumber
\end{eqnarray}
The LAEMW-powered \gr\ emission stage shows a characteristic initial
phase during which the luminosity remains nearly constant. This phase is
expected to be followed by a phase in which the luminosity decreases as
$\propto t^{-2}$. As shown in Blackman and Yi (1998), 
the \gr\ emission
via the synchrotron-Compton process
gives the luminosity of the peak $\propto t^{-1}$.
We note that if the proton fraction changes after some complicated
acceleration processes, then the synchrotron-Compton emission process and
the above scalings could be significantly different. 

To get \grs\ directly from the LAEMW, there must be a density in the 
environment that is below the Goldreich-Julian density so
the currents cannot be supported in the plasma and a
pair cloud is spontaneously generated.  The Goldreich-Julian
density scales with the ambient magnetic field.  If that
field falls off like $R^{-1}$ in the propagating Poynting flux, 
then beyond the helium core
at R$\gta10^{6}$~km, the Goldreich-Julian density will
be (equation \ref{GJdensity})
$\rho\lta 10^{-11}$~\gcm3.  Densities this low might
occur immediately beyond the star (and the jet) if the star is embedded
only in the interstellar medium, but it is likely that
the progenitor of a Type Ib/c supernova is surrounded by a wind or
other mass resulting from mass-loss processes.  For
a constant velocity wind at $10^8~v_8$~\cms\ carrying
mass at a rate $10^{-5}{\dot M_{-5}}$~\m~yr$^{-1}$,
the density is   
\begin{equation}
\label{winddensity}  
\rho = 5\times10^{-9}~{\rm g~cm^{-3}}~{\dot M_{-5}}v_8^{-1}R_{5}^{-2},
\end{equation}
with radius in units of $10^5$ km.
This density will be less than the Goldreich-Julian density
for R$\gta10^{6}$~km, several times the radius of the
helium core.

Although the possibility
of \gr\ emission by pair cascade is not ruled out, 
it is important to consider how the UMHDW/LAEMW escape from the
stellar core and associated processes.
A sub-relativistic matter-dominated MHD jet leaves behind a baryon-rich
environment. The common statement is that if the baryon loading
is too high a \grb\ of observed properties cannot be produced.
We show below that this is not necessarily the case.

Since the plasma density (baryons, leptons, and photons) 
is high, the perfect MHD condition
is always maintained throughout the phase when the UMHDW
propagate within the stellar core. Due to the high radiation density 
and the large amount of high temperature stellar material, 
the optical depth for outgoing radiation is exceedingly high. 
This makes the \gr\ emission efficiency very low. 
Under these circumstances, the bulk of the energy supplied by the
UMHDW may be used to accelerate shocks parallel to the 
rotation axis and the axis of the matter jet. 
After the shock breaks out, \gr\ emission could occur as the kinetic
energy of the shock is dissipated into particle acceleration 
and subsequent synchrotron-Compton emission. We note one 
important difference between this situation and that of the original
Colgate mechanism.  As noted in \S 7.1, in the Colgate mechanism 
there is an
impulsive input of energy in the stellar core and a single shock
that propagates outward.  This leads to a single, short pulse
of hard emission.  This process may, indeed, work when the 
shock driven by the UMHDW first encounters the steep density
gradient at the stellar surface or at the tip of the jet, 
but in the current situation
the pulsar and the UMHDW/LAEMW continue to drive shocks
for an extended time. The density profile will be
altered as the continuing shock energy is deposited and that
reaction must be considered self-consistently.

It is difficult to estimate the \gr\ emission efficiency in this process. 
Even for a very low efficiency 
$\lta 10^{-4}$, however, a \grb\ event such as 
SN~1998bw/GRB980425 could be generated provided that the UMHDW power is 
large enough, $\gta 10^{51}$~\erg-s\ as has been assumed. 
If the efficiency of production of \grs\ approaches unity, 
then, with appropriate collimation, a \grb\ detectable at
cosmological distances could be produced.
Aside from the complex
details of the \gr\ production, the overall
bolometric luminosity evolution is expected to follow the time evolution
described by equation (\ref{Lobs2}).

In this context, we take note of circumstances when the deposition
of energy into baryons does not necessarily doom the production
of a \grb.  As noted by Protheroe \& Bednarek (1999), if protons
can be accelerated to energies where pion production is efficient,
then \grs\ can be produced in the subsequent pion decay.  To
reach this regime, the protons must be given at least mildly
relativistic energies, in excess of 1 GeV. 
This requires that the energy of the UMHDW/LAEMW be shared with
less than 0.005 \m $E_{52}$ of baryons.  
The shock driven by the UMHDW/LAEMW should boost some protons to
even higher bulk velocities and the energy deposited by
the UMHDW/LAEMW will plausibly go into a mass less than the total mass of the 
precursor subrelativistic jet
if their effects are concentrated in the relatively low
density matter along the axis of the jet or the matter
immediately surrounding the jet.  Another alternative
is that the protons will be accelerated by the 
momentum associated with the Poynting flux.  Deposition of
the energy of the pulsar spindown via UMHDW selectively
into the bulk motion of protons with energies substantially
above the pion production threshhold is thus a distinct
possibility.  Thus rather than being a handicap, the
``baryon loading" of the jet could be an advantage in
the conversion of pulsar energy into \gr\ energy.  

Although a quantitative calculation is required, we envisage 
a process by which the protons are accelerated to high
energy with the associated efficient production of pions.
To produce the pions, the protons must collide with a ``target."
In the present context, a good candidate for the target is
the wind expected to be present around the progenitor.
The stopping length for proton-proton interaction to
produce pions is about 100 cm$^2$ gm$^{-1}$. 
For the density given by equation (\ref{winddensity}),  
the column depth of the wind is
\begin{equation}
\label{columndepth}
l = 50~{\rm~g~cm^{-2}}~{\dot M_{-5}}v_8^{-1}R_{5}^{-1},
\end{equation}    
with radius again in units of $10^5$ km.
Thus a substantial wind might provide the stopping medium for
the conversion of high-velocity protons to pions
in the vicinity of the surface of a hydrogen-stripped star. 

The pions produced when the high-energy protons collide
with the wind then decay and produce very energetic \grs.
Note that the stopping length for \grs\ is about 30 cm$^2$ gm$^{-1}$,
so a ``target" that stops protons will be somewhat optically
thick to \grs.  
These pion-decay high energy \grs\ do not immediately correspond
to the observed \grs\ in a \grb. Rather, they induce
a pair cascade through photon-photon collision. 
This pair fireball could then produce the observed \grb.
The strength of a \grb\ produced in this way will
depend on the relative efficiency of production of
pions versus neutrinos.  The latter would be an energy
sink for this process.

Note that equation (\ref{columndepth}) neglects any affect of
the ambient or stellar magnetic fields or those associated with
the UMHDW/LAEMW.  If the field were strong and the interparticle spacing
small, the shock generated by the UMHDW/LAEMW could be collisionless
and there might be no pion production.  Alternatively,
if the shock accelerates protons into a small magnetic
field then this estimate would be valid.  As an example, if
the field scales as $R^{-2}$ from a central dipole of
$10^{16}$ G, then the Larmor radius near the surface of the
star would be about $10^{-4}$ cm.  If the field is associated with
the LAEMW and scales as $R^{-1}$, then the Larmor radius
would be about $10^{-8}$ cm.  At the density given in the wind
by equation (\ref{winddensity}), the interparticle spacing is
about $10^{-5}$ cm.  Thus gyrating protons might still collide 
in the $10^8$ G dipole field with an effective
path length enhanced by the gyration, and equation (\ref{columndepth})
could be a lower limit, or they could be rendered collisionless
in the $10^{12}$ G field associated with the LAEMW.  On the
other hand, since the UMHDW/LAEMW could dominate the dynamics of
the protons, the idea of a shock among particles carried along
by the LAEMW might be moot anyway (Usov 1999).  Clearly, the
physics associated with these processes is complex and in
need of deeper understanding.  The only point to be made here
is that the physics of proton acceleration and pion production
in the wind might be relevant and needs further study. 
 
\section{Disscussion and Conclusions}

We have shown that the contraction phase of a proto-neutron star
could result in a substantial change in the physical properties
of the environment.
The following parameters are relevant:

\begin{tabbing}
XXXXXXXXXXXXXXXXXXXXXXXXXXX\=\kill
{$\bullet$} $P_{PNS}\simeq 25 ms$ \> {$\bullet$} $P_{NS}\simeq 1 ms$\\
{$\bullet$}  $R_{PNS}\simeq50$ km \> {$\bullet$} $R_{NS}\simeq$10~km\\
{$\bullet$}  $R_{A,PNS}\approx150$ km \> {$\bullet$} $R_{A,NS}\approx$150~km\\
{$\bullet$} $R_{LC}\simeq10^3$~km \>{$\bullet$} $R_{LC}\simeq$50~km\\   
{$\bullet$} $I_{PNS}=kR^2_{PNS}M_{NS}\simeq3\times10^{46}$ c.g.s. \>
{$\bullet$} $I_{NS}= kR^2_{NS}M_{NS}\simeq10^{45}$ c.g.s.\\
{$\bullet$} $\tau_{\rm conv}\simeq 1 {\rm ms}$ \> 
{$\bullet$} $\tau_{\rm conv}\simeq 1 {\rm ms}$ \\  
{$\bullet$} $R_0\sim P_{PNS}/\tau_{\rm conv}\gg1$ \> {$\bullet$} $R_0\gta1$ \\ 
\ \ \ $\rightarrow$ no dynamo \> \ \ \  $\rightarrow$ 
$\alpha - \Omega$ dynamo \\
{$\bullet$} magnetic field amplified within \> 
{$\bullet$} exponential growth of field\\
\ \ \ $\sim$10s by linear amplification or \> \ \ \ \\ 
\ \ \ flux-freezing \>\ \ \ \\  
{$\bullet$} MHD jet \& torque action \> {$\bullet$} UMHDW generated: 
 ${\vec S}=\frac{c}{4\pi}({\vec E}\times{\vec B})$ \\
{$\bullet$} hole punched \& envelope torqued \> 
{$\bullet$} relativistic jet \\
\ \ \ \> \ \ \ envelope accelerated and heated by waves\\
\end{tabbing}
\vskip8mm

When the rotating magnetized neutron star first forms there is 
likely to be linear amplification of the magnetic field and
the creation of a matter-dominated jet, perhaps catalyzed by 
subrelativistic MHD effects, 
up the rotation axis.  The energy
of the proto-neutron star is sufficient to power a significant
matter jet, but unlikely to generate a strong \grb.  The matter jet could
generate a smaller \grb\ as seems to be associated with
SN~1998bw and GRB~980425 by the Colgate mechanism as
it emerges and drives a shock down the stellar density gradient
in the absence of a hydrogen envelope, e.g., in a Type Ib/c
supernova.  As the neutron star cools, contracts,
and speeds up, two significant things happen.  One is that
the rotational energy increases.  The energy becomes significantly
larger than required to produce a supernova and sufficient, 
in principle, to drive a cosmic \grb\ if the collimation
is tight enough and losses are small enough.  
In addition, the light cylinder contracts
significantly, so that a stationary dipole field cannot form
and the emission of strong UMHDW occur.  Tight collimation 
of the original matter jet and of the subsequent flow of UMHDW
in a radiation-dominated jet is expected.

The UMHDW will propagate as intense low frequency, long
wavelength Poynting flux.  They may be isotropized to act
as a relativistic fluid, but not thermalized since
they have a frequency much less than the plasma frequency.
The UMHDW ``bubble" could be strongly Rayleigh-Tayor unstable,
but still may propagate selectively with small opening angle
up the rotation axis as an UMHDW jet.  Alternatively, the
impulsive production of UMHDW could render the stellar matter
nearly irrelevant as a confining medium.
If a UMHDW jet forms, it can drive shocks which may selectively propagate 
down the axis of the initial matter jet or around the perimeter of
the matter jet.  The shocks associated with the UMHDW jet could generate
\grs\ by the Colgate mechanism as they propagate down
the density gradient at the tip of the jet or there
could be bulk acceleration of protons to above
the pion production threshold.  The protons could produce
copious pions upon collision with the surrounding wind,
thus triggering a cascade of high energy \grs, pairs,
and lower-energy \grs\ in an observable \grb.
Yet another alternative is that the UMHDW  
could eventually propagate into such a low density
enviroment that they directly induce LAEMW and pair cascade
(see also Thompson \& Madau 1999).

The radiation-dominated jet cannot form for several seconds as
the neutron star contracts, spins up, and generates a large
magnetic field, but then it propagates faster than the matter-
dominated jet.  In this circumstance, the matter jet could precede 
or follow the radiation-dominated jet.  In the former case an 
X-ray precursor could be generated.  In the latter case the matter
jet might not be conspicuous at all.            

There are a number of reasons why the processes we have
outlined may not be as effective as we have assumed.
One is that the rotation of the neutron star may
prevent contraction to the high densities and rotation
rates on the timescales we have assumed 
(Fryer \& Heger 1999).  This could affect the
convection in the neutron star and hence the generation of 
the magnetic field by the $\alpha-\Omega$ dynamo mechanism.   
This is a complex issue, of course,
since the presence of the magnetic field will lead to 
energy loss and conditions of greater contraction,
as we have invoked here (Symbalisty 1984).

Another complication we have ignored is that the 
mean dipole field that forms may have its axis tilted
with respect to the spin axis.  This may not be a
critical factor, since the subsequent dynamics may be
dominated by the density distribution surrounding
the neutron star that is predominantly set by
the angular momentum, not the magnetic field.
The question of what fraction of the pulsar energy
goes to drive quasi-spherical expansion and what
fraction propagates as co-linear UMHDW clearly
requires greater study.  We also noted that if 
the UMHDW are thermalized in the stellar core the
result will be the copious production of electron/positron
pairs.  At first, such a pair cloud will behave like
a relativistic fluid so there may be little difference
in the dynamics.  As Rayleigh-Taylor instabilities ensue,
the pairs might get mixed with ordinary matter and the
positrons annihilate.  Even if the pair bubble escapes
up the axis, as we envisage for non-thermalized UMHDW,
it will expand as it propagates, especially after 
breaking through the stellar surface.  Much of the
thermal energy could then be lost to kinetic energy
by adiabatic expansion. If there is no ``working surface"
with which this pair cloud could collide, then much
of the energy could be lost.  On the other hand,
the sort of wind expected (equation \ref{winddensity}) would
give a stopping length (equation \ref{columndepth}) that
will easily stop the pairs.  The issue would then
be the efficiency of conversion of their kinetic
and rest mass energy into \grs.  

There is an interesting question of the opposite sort
concerning the possibility that the production of large
energy in UMHDW could be too effective.  For instance,
if an UMHDW jet could propagate through the naked
core of a Type Ic, but would get stopped and dissipated
in a red supergiant, then we might find that the
explosion energy of Type II is systematically higher
than that of Type Ic.  The fact that there is no
clear evidence for this may suggest that the production
of $10^{52}$ ergs in UMHDW is not a common occurrence in
core collapse, but this question requires further study. 

The question of
whether or not pulsar spin-down will produce a \grb\ 
depends on such factors as the initial rotation
rate, the strength of the dipole field that evolves,
the tilt of such a field compared to the spin axis,
and the density of the progenitor wind.   Issues of uncertain
physics aside, it is clear that this mechanism might
not be robust in the production of \grbs, but might
produce \grbs\ of varying strength depending on 
natural variation in the circumstances of a given
collapse event.

Any \grs\ emitted by any of these processs 
are likely to be strongly collimated.
The luminosity of the emitted radiation will depend on the
geometry of that emission.  We have noted here that the energy
produced by the spin-down of the pulsar could emerge from
the stellar surface along the axis of a low-density matter jet, or in 
an annulus surrounding a high density jet.  Either of 
these cases will give a Lorentz factor that depends strongly
on the aspect angle of the observer.  Computation of the 
resulting luminosity is thus distinctly non-trivial.

Simple \grb\ models invoking
collimation assume that the energy in a collimated burst 
scales simply as $\Delta\Omega/4\pi$ compared to that deduced
in an isotropic geometry.  This assumes that the collimated
jet nevertheless expands as the section of a sphere with 
a cross section scaling as $R^2$.  We note that if the 
relativistic flow is truly collimated, this may yet be an
overestimate of the energy required to power a given observed
\grb.  The calculation of a jet emerging from a stellar core
by Khokhlov et al. (1999) shows that the dynamical jet that
precedes any possible \grb\ is nearly linearly collimated and
does not expand in cross section as $R^2$.  In the absence
of an understanding of the dynamics of the subsequent flow
of UMHDW/LAEMW, we do not know the geometry of the relativistic
flow, but if the cross sectional area grows less strongly
than $R^2$, then the solid angle $\Delta\Omega/4\pi$ will
be a {\it decreasing} function of distance.  
The energetics (and hence the
derived rates of occurence) of \grbs\ will depend not only
on the fact, but the geometry of the collimation.  If the
cross section expands less rapidly than $R^2$, then substantially
less energy may be required to produce a given observed \grb.

The observed nature of any \grb\ will depend on whether or not one
is directly witnessing a strongly relativistic bulk flow.  There is
general agreement that the observed afterglows of the 
cosmic \grbs\ represent relativistic blast waves.  The
question of whether that is true or not for the primary
\grb\ is still in contention with models based on 
internal shocks in a ``central machine," 
in which the burst duration is the fundamental physical time scale 
of the energy production process (e.g. Piran 1999, and references
therein), vying against models invoking relativistic flows and 
external shocks for which there is strong Lorentz contraction
of the timescales between the production mechanism and the observer
(e.g. Dermer, B\"ottcher \& Chiang 1999).  
Fenimore \& Ruiz (1999) have
recently argued that a central machine is favored (see also
Heinz \& Begelman 1999).  

One of the implications of this uncertainty is the location of the
\grb.  If an external blast wave is involved, then  
a \grb\ with a time scale in the observer frame of $t_{obs}$  
has a propagation distance of $\Gamma^2 c t_{obs}$. 
For a \grb\ of 10 s this distance is about 
$3\times 10^{15}$ cm for a Lorenz factor of $\Gamma \simeq 100$.
This is much larger than the radius of the hydrogen-deficient
stellar core we
are considering here, $R_{core}\simeq 10^{10}$ cm.  It is
not clear why the energy produced by shocks and UMHDW/LAEMW would
be dumped at a radius as large as $10^{15}$ cm.    Shocks should
deposit their energy as they emerge from the star, and
the density falls below the Goldreich-Julian density at 
only $10^{11}$ cm even for a relatively dense wind.  If 
relativistic protons are generated by shocks or other
mechanisms, they could also plausibly be stopped near
the stellar surface by a substantial wind.  Thus for all the
mechanisms we sketch here, the most logical site of the
production of any \grb\ is near the stellar surface.  This
implies that if this general process of pulsar spindown from
massive star core collapse has any role in producing
\grbs\ it will most plausibly serve as a ``central engine."
We note that in this case, the natural time scale for
any \grb\ is about 5 to 10 sec, the cooling, spin down time
for the neutron star.  
On the other hand, the eruption
of shocks and UMHDW/LAEMW from the stellar surface may occur in
a region small compared to the stellar surface, so
considerably shorter timescales might be manifest for
central burst peaks. If the emission from the pulsar
is prolonged, then one might also witness time scales
associated with the processes of production of the
UMHDW, for instance various instabilities, as well.
The shortest time scales of any substructure
would be $\simeq 10^{11}~{\rm cm}/\Gamma^2c$ or about 
$0.3$ ms for $\Gamma = 100$.
 
The deposition of a large amount of energy at the stellar
surface could, of course, result in a subsequent 
relativistic blast wave and associated afterglow.   
The processes we are discussing would produce a maximum
``isotropic equivalent" energy of $4\pi E/\Delta\Omega\simeq
10^{54}~{\rm erg}$ for $E_{52}\simeq 1$ and $\Delta\Omega/4\pi
\simeq~0.01$.  The external mass required to decelerate this energy, 
$E\Gamma^{-2}c^{-2}$,  requires a spherically-distributed mass of
$\simeq 5\times10^{-5}~E_{52}$~\m\ for $\Gamma = 100$.
The mass in the wind out to a radius R is
\begin{equation}
\label{windmass}
M_{wind} = 3\times10^{-6}~\m\ {\dot M_{-5}}v_8^{-1}R_{15},
\end{equation}
with $R$ in units of $10^{15}$ cm.
Thus energy emitted by the processes we have outlined
will not be decelerated 
until a radius of about $10^{16}$ cm.  Even ignoring issues
of entrainment and asymmetries in the wind density profile,
this radius is sensitive to the energy of the burst, the
degree of collimation, the value of the Lorentz factor
and the parameters of the wind.

The point of this paper is not to establish that core collapse
and pulsar formation will lead to \grbs, but to establish
that this environment gives a framework in which to 
quantitatively address questions of physics that are
germane to the nature of the core collapse process and to 
potential \gr\ production.  It seems very clear that rotation
and magnetic fields have a strong potential to create
axial matter-dominated jets that will drive strongly asymmetric
explosions for which there is already ample observational
evidence in Type II and Type Ib/c supernovae, their remants,
and in the pulsar velocity distribution. 
The potential to also create strong flows of UMHDW/LAEMW
serves to reinforce the possibility to generate asymmetric explosions. 
These asymmetries will affect nucleosynthesis and issues such
as fall-back that determine the final outcome to leave behind
neutron stars or black holes. In addition, the presence of 
matter-dominated and radiation-dominated jets might lead
to bursts of \grs\ of various strengths.  The issue of the
nature of the birth of a ``magnetar" in a supernova explosion
is of great interest independent of any connection to \grbs.
Highly magnetized neutron stars might represent one out of
ten pulsar births.  Production of a strong \grb\ might be
even more rare.

The authors are grateful to Rob Duncan, Alexei Khokhlov, Elaine Oran ,
and Martin Rees for helpful discussions.  This research was supported 
in part by NSF Grant 95-28110, NASA Grant NAG 5-2888, a grant 
from the Texas Advanced Research Program, KRF 1998-001-D00365 (to IY) 
and the Ewha University Faculty Research Fund (to IY).


\begin{references}


Asseo, E., Kennel, C. F., \& Pellat, R. A. 1978, A\&A, 65, 401

Baring, M. G. \& Harding, A. K. 1997, ApJ, 491, 663

Bisnovatyi-Kogan, G. S. 1971, Soviet Astronomy AJ, 14, 652

Bisnovatyi-Kogan, G. S., Popov, Iu. P. \& Samokhin, A. A. 1976, 
ApSS, 41, 287

Blackman, E. G. \& Yi, I. 1998, ApJ, 498, L31

Blackman, E. G., Yi, I., \& Field, G. B. 1995, ApJ, 473, L79

Burrows, A., Hayes, J. \&  Fryxell, B. A. 1995, ApJ, 450, 830 

Burrows, A. \& Lattimer, J. M. 1986, ApJ. 307, 178 

Bloom, J. S. et al. 1999, Nature, 401, 453


Cen, R. 1998, ApJ, 507, L131

Chevalier, R. A. \& Li, Z.-Y. 1999a, ApJ, 520, L29

Chevalier, R. A. \& Li, Z.-Y. 1999b, ApJ, submitted, astro-ph/9908272

Colgate, S. A. 1974, ApJ, 187, 333

Colgate, S. A. 1975, ApJ, 198, 439


Danziger, J. et al. 1999, in
The Largest Explosions Since the Big Bang: Supernovae and Gamma-Ray
Bursts, eds. M. Livio, K. Sahu \& N. Panagia, in press

Dermer, C. D., B\"ottcher, M. \& Chiang, J. 1999, ApJ, 515, L49

Duncan, R. C. \& Thompson, C. 1992, ApJ, 392, L9

Fenimore, E. E. \& Ramirez-Ruiz, E. 1999, PASP Conf. Proc. Gamma-Ray Bursts:
The First Three Minutes (Graftavallen Sweden), submitted, astro-ph/9906125

Fesen, R. A. \& Gunderson, K. S. 1996, ApJ, 470, 967
 
Fryer, C. L. \& Heger, A., 1999, ApJ submitted, astro-ph/9907433 

Galama, T. J. et al. 1998, Nature, 395, 670 

Galama, T. J. et al. 1999, ApJ, submitted, astro-ph/9907264

Germany, L. M., Reiss, D. J., Sadler, E. S., Schmidt, B. P. \& Stubbs, C. W.
1999, ApJ, submitted, astro-ph/9906096

Goldreich, P. \& Julian, W. H. 1969, ApJ, 160, 971

Harrison, F. A. et al. 1999, ApJ, 523, L121 


Heinz, S. \& Begelman, M. C. 1999, ApJ, submitted, astro-ph/9908026

M\"uller, E. \& Hillebrandt, W. 1979, A\&A, 80, 147

H\"oflich P. 1991, A\&A, 246, 481   

H\"oflich, P., Wheeler, J. C., \& Wang, L. 1999, ApJ, 521, 179  

Iwamoto et al. 1998, Nature, 395, 672

Jeffery  D.J., 1991, ApJ, 375, 264  
                                            
Khokhlov  A.M., H\"oflich P. A., Oran E. S., Wheeler J.C.
Wang, L, \& Chtchelkanova, A. Yu. 1999, ApJ, 524, L107 

Klu\'zniak, W. \& Ruderman, M. 1998, ApJ, 505, L113 

Kouveliotou, C., Strohmayer, T., Hurley, K., Van Paradijs, J., 
Finger, M. H., Dieters, S., Woods, P., Thompson, C. \& Duncan, R. C.
1998, ApJ, 510, 115 

Krolik, J.H. \& Pier, E. A. 1991, ApJ, 373, 277   


Kulkarni et al. 1999, Nature, in press, astro-ph/9902272 

Lamb, D. Q. 1999, A\&A, in press.

LeBlanc, J. M. \& Wilson, J. R. 1970, ApJ, 161, 541

Leonard, D. C., Filippenko, A. V., Barth, A. J. \& Matheson, T. 1999,
ApJ, submitted, astro-ph/9908040

Lucy, L.B. 1988, Proc. of the 4th George Mason conference,
 ed. by M. Kafatos, Cambridge University Press, p. 323
                                                         
MacFadyen, A. \& Woosley, S. E. 1999, ApJ, 524, 262

Matzner, C. D. \& McKee, C. F. 1999, ApJ, 510, 403

Meier, D., Epstein, R. I., Arnett, W. D. \& Schramm, D. N. 1976,
ApJ, 204, 869

M\'endez M., Clocchiatti A., Benvenuto  G., Feinstein  C., Marraco  U.G.,
1988, ApJ, 334. 295
                      
Mezzacappa, A. et al. 1998, ApJ, 495, 911.

Michael, C. 1984, ApJ, 284, 384

Nagataki, S. 1999, ApJ, 511, 341 

Nakamura, T. 1998, Prog. Theor. Phys. 100, 921

Ostriker, J. P. \& Gunn, J. E. 1971, ApJ, 164, L95

Paczy\'nski, B. 1986, ApJ, 308, L43

Paczy\'nski, B. 1991, Acta Astron, 41, 257

Paczy\'nski, B. 1998, ApJ, 494, L45

Piran, T. 1999, Physics Reports, 314, 575

Protheroe, R. J. \& Bednarek, W. 1999, astro-ph/9904279

Reed, J, Hester, J. \& Winkler, F. ApJ, 1999, submitted.

Rees, M. J. \& M\'esz\'aros, P. 1992, MNRAS, 258, 41p
 
Reichart, D. E. 1999, ApJ, submitted, astro-ph/9906079
 
Rhoads, J. E. 1997, ApJ, 487, L1

Rhoads, J. E. 1999, ApJ, 525, 737 


Sari, R. Piran, T. \& Halpern, J. P. 1999, ApJ, 519, L17

Stanek, K. Z. et al. 1999, ApJ, 522, L39  

Strom R., Johnston H.M., Verbunt F. \& Aschenbach B. 1995, Nature, 373, 587

Sunyaev et al. 1987, Nature, 330, 227

Symbalisty, E. M. D. 1984, ApJ, 285, 729

Taylor J.H., Manchester R.N. \& Lyne A.G. 1993, ApJS, 88, 529
                                                                 
Trammell,  S.R., Hines, D.C. \& Wheeler, J.C. 1993, ApJ, 414, L21   

Tran H.D., Filippenko A.V., Schmidt G.D., Bjorkman K.S., Januzzi B.J., 
Smith P.S.  1997, PASP, 109, 489
                           
Thompson, C. 1994, MNRAS, 270, 480

Thompson, C. \& Duncan, R. 1993, ApJ, 408, 194

Thompson, C., Duncan, R. C., Woods, P. M., Kouveliotou, C.,
Finger, M.H. \$ van Paradijs, J. 1999, ApJ, submitted, astro-ph/9908086 

Thompson, C. \& Madau, P. 1999, ApJ, submitted, astro-ph/9909111 

Tueller J., Barthelmy, S., Gehrels, N., Leventhal, M., MacCallum, C.J., 
Teegarden, B. J.  1991, in: 
Supernovae, ed. S.E. Woosley, Springer Press, p. 278
                                                                    
Usov, V. V. 1992, Nature, 357, 452

Usov, V. V. 1994, MNRAS, 267, 1035

Usov, V. V. 1999, astro-ph/9909435

Wang L.,  Wheeler J.C., 1998, ApJ, 584, L87

Wang, L., Wheeler, J. C. \& H\"oflich, P. 1997, in SN 1987A: Ten Years After
ed. M. M. Phillips and N. Suntzeff (Provo: Ast. Soc. of the Pacific),
in press

Wang, L., Wheeler, J. C. \& H\"oflich, P. 2000, ApJ, submitted

Wang, L., Wheeler, J. C., Li, Z. W., \& Clocchiatti, A. 1996, ApJ, 467, 435
      
Wheeler, J. C. 1999, in 
The Largest Explosions Since the Big Bang: Supernovae and Gamma-Ray
Bursts, eds. M. Livio, K. Sahu \& N. Panagia, in press    

Woosley, S. E. 1993, ApJ, 405, 273

Woosley S., Eastman R. \& Schmidt M. 1998, ApJ, 516, 788

\end{references}
\end{document}